\documentclass[aip,pop,sd,reprint,floatfix]{revtex4-1}

\usepackage{amsmath,amssymb,amsfonts,amstext,graphics,graphicx,subfigure}
\usepackage[colorlinks]{hyperref}
\usepackage{color}

\usepackage{mathrsfs}

\def\Xint#1{\mathchoice
{\XXint\displaystyle\textstyle{#1}}%
{\XXint\textstyle\scriptstyle{#1}}%
{\XXint\scriptstyle\scriptscriptstyle{#1}}%
{\XXint\scriptscriptstyle\scriptscriptstyle{#1}}%
\!\!\int}
\def\XXint#1#2#3{{\setbox0=\hbox{$#1{#2#3}{\int}$ }
\vcenter{\hbox{$#2#3$ }}\kern-.5\wd0}}

\def\dashint{\Xint-}

\def\cala{\mathcal{A}}

\def\calc{\mathcal{C}}
\def\cald{\mathcal{D}}

\def\calf{\mathcal{F}}

\def\calo{\mathcal{O}}

\def\cals{\mathcal{S}}

\def\R{\mathbb{R}}

\def\ppt{\frac{\partial}{\partial t}}

\def\bq{\begin{equation}}
\def\eq{\end{equation}}
\def\bqy{\begin{eqnarray}}
\def\eqy{\end{eqnarray}}

\def\al{\alpha}

\def\de{\delta}

\def\ep{\epsilon}

\def\et{\eta}

\def\la{\lambda}

\def\om{\omega}

\def\varep{\varepsilon}

\def\p{\partial}

\def\and{\quad\mathrm{and}\quad}

\def\ncr{\nonumber\\}

\newcommand{\refeq}  [1] {(\ref{#1})}

\newcommand{\refFig} [1] {Figure~\ref{#1}}

\newcommand{\initcond}[1]{\overset{\circ}{#1}}
\newcommand{\define}{:=}

\newcommand{\Half}{\frac{1}{2}}

\begin{document}

\title{An integral transform technique for kinetic systems with collisions}
\author{J.~M.~Heninger}
\email{jeffrey.heninger@yahoo.com}
\affiliation{Department of Physics and Institute for Fusion Studies, 
The University of Texas at Austin, Austin, TX, 78712, USA}
\author{P.~J.~Morrison}
\email{morrison@physics.utexas.edu}
\affiliation{Department of Physics and Institute for Fusion Studies, 
The University of Texas at Austin, Austin, TX, 78712, USA}
\date{\today}

\begin{abstract}
 
The  linearized Vlasov-Poisson system can be exactly
solved using the {\it $G$-transform},  an integral 
transform introduced in Refs.~\onlinecite{MorrPfir92,pjm94,Morr00} that  removes 
the electric field term, leaving a simple advection equation. We investigate
how this integral transform interacts with the Fokker-Planck collision
operator. The commutator of this collision operator with the $G$-transform
(the ``shielding term'') is shown to be negligible. We exactly solve 
the advection-diffusion equation without the shielding term. This
solution determines when collisions dominate and when advection 
(i.e.\  Landau damping) dominates. This integral transform can also be
used to simplify gyro-/drift-kinetic equations. 
We present new gyrofluid
equations formed by taking moments of the $G$-transformed equation.
Since many gyro-/drift-kinetic codes use Hermite polynomials as basis elements,
we include an explicit calculation of their $G$-transform. 
%
\end{abstract}

\maketitle



\section{Introduction}
\label{sec:intro}
 
	One of the intriguing features of the Vlasov-Poisson system is 
	the ability	of an electrostatic wave to damp, even in the absence 
	of any dissipation mechanism.\cite{Lan46} This can occur because 
	the electrostatic wave couples to a continuum of modes describing 
	all of the possible ways that the distribution function can vary
	in velocity space. 
	
	The linearized Vlasov-Poisson system can be solved using the  Laplace 
	transform,\cite{Lan46,LenBern58} normal modes,\cite{VanKampen55}  a Green's function,\cite{WBT62,Call14} or
	numerically by a variety of codes (e.g.\ Refs.~\onlinecite{CK76,pjmHGM12,PezVal16,pjmKKS17}).
	Another method for describing Landau damping, which employs an integral transform
	in velocity space based on the Hilbert transform,  was introduced in 
	Refs.~\onlinecite{MorrPfir92,pjm94,Morr00}. Applying this {\it  $G$-transform}  
	to the linearized Vlasov-Poisson equation 
	completely removes the electric field term. The resulting
	advection equation is trivially solved, and then the inverse
	transform is applied to return the solution to the original
	coordinates.	
	
	Versions of this integral transform exist for a variety of
	models,\cite{pjm03} and there is one that can be used
	effectively on any collisionless kinetic model with one
	velocity dimension. 
	
	A simple and important way to extend the linearized Vlasov-Poisson
	system is to add collisions. The Landau-Boltzmann collision
	operator for Coulomb collisions is usually unmanageable, so 
	simpler collision operators are often used instead. A good collision
	operator, such as the Fokker-Planck 
	operator,\cite{Ray91,Fok14,Chan43,LenBern58,DegDes92} 
	acts as a diffusion in velocity space, conserves particle number, 
	and has a Maxwellian equilibrium. Typically, the collision
	frequency is assumed to be small. Even if the collision frequency 
	is small, collisions shouldn't be completely neglected. 
	Landau damping produces fine structures in velocity space 
	for which the diffusion eventually becomes important. 
	
	The picture of Landau damping is similar to Kolmogorov turbulence: 
	energy which is input on large velocity scales cascades down 
	to smaller velocity scales until it reaches a dissipation 
	velocity scale where the collisions are important.
	There is also an intermediate velocity scale where collisions 
	are important for particles with small parallel velocities.\cite{WhiHaz17} 
	Since the damping rate typically depends 
	more on the creation of fine structures in velocity space than on 
	the details of the collision operator, replacing the
	complete collision operator with a simplified version is reasonable.
	Landau damping causes the effective dissipation rate to be much 
	higher than the collision frequency. The importance of the 
	fine structures in determining the dissipation rate can be seen 
	by noticing that if the initial conditions have fine structures in
	velocity space, then the damping rate can be much higher than if 
	the initial conditions differ from a Maxwellian in a more gradual 
	way.\cite{PezVal16}
	
	Gyrokinetics and drift-kinetics are a natural places to look for applications for the 
	$G$-transform because they have only one velocity dimension.
	In many applications, the plasma has a strong magnetic field. 
	A particle's motion perpendicular to this field mostly follows small
	circles around the field lines. This motion does not need to be
	resolved. Instead, the three-dimensional Vlasov-Poisson equations are
	integrated over the perpendicular velocity dimensions. Drift 
	velocities which depend on the fields can be added to preserve the
	important aspects of the perpendicular motion. The resulting
	system has three spatial dimensions, one velocity dimension and time.
	\cite{FriChen82,BrizHa07,Krom12}
	
	The $G$-transform has been used to analyze data from laboratory
	experiments.\cite{Skiff96,Skiff00,Skiff02,Skiff02b} 
	Doppler-resolved laser induced fluorescence measures the 
	dependence of the distribution function on the velocity 
	parallel to the laser beam. $G$-transforming the data at a single
	point yields the amplitudes of the van Kampen 
	modes excited by an externally created wave. The spatial 
	dependence of the electric field can be determined from the
	van Kampen mode amplitudes using a measurement of the velocity 
	distribution function at a single point.
	The $G$-transform could also be used on satellite data to measure
	Landau damping and nonlinear energy transfer between fields 
	and particles in the heliosphere.\cite{Howes16,Howes18}
	
	The paper is organized as follows. 
	In Sec.~\ref{sec:GVlasov} we review general properties of
	the $G$-transform and its use on the  one-dimensional 
	linearized Vlasov-Poisson system. This is followed by
	Sec.~\ref{sec:Collisions} that deals with collisions. 
	Here we show that the $G$-transform almost commutes with 
	the collision operator. The resulting equations have an 
	exact solution which allows us to easily compare advection
	and diffusion. These results are robust to changes in 
	initial conditions or the collision operator. 
	Sec.~\ref{sec:Gyrokinetics} shows how the $G$-transform 
	can also be used to simplify gyro-/dirft-kinetic equations. 
	The resulting equations, which no longer have a parallel 
	electric field, can be solved numerically using a basis
	of Hermite polynomials or we can take moments of them to 
	get novel drift-fluid equations. 
	We conclude in Sec.~\ref{sec:Conclusion}.
	
\section{One-Dimensional Vlasov-Poisson}
\label{sec:GVlasov}

	One of the simplest models for a plasma is the one-dimensional
	Vlasov equation for a single  species  plasma, 
	\bq
	 \frac{\p f}{\p t} + v \frac{\p f}{\p x} - \frac{e}{m}
		\frac{\p \phi}{\p x} \frac{\p f}{\p v} = 0\,, 
		\label{vlasov}
		\eq
where the electric field is determined self-consistently using the Poisson equation,  
\bq
\nabla^2 \phi =  4\pi e \int_\R f dv - 4 \pi \rho_b\,,
\label{poisson}
\eq
with $e$ and $m$ being the electron charge and mass, respectively,  and $\rho_b$ being a neutralizing background positive charge density.  Linearizing \eqref{vlasov} and \eqref{poisson} about an arbitrary homogeneous equilibrium, $n_0 f_0(v)$ and  Fourier transforming  the perturbation in position, yields 
	\bq
		\frac{\p f_k}{\p t} + i k v \,f_k - i \,\frac{\om_P^2}{k}
		\int_\R f_k \,dv' \,\frac{d f_0}{d v} = 0,
		\label{OrigVlasov}	
	\eq
	which we consider for an arbitrary initial conditions $\initcond{f_k}(v)$.
	In \eqref{OrigVlasov}	  the third term comes from substituting the linearized Poisson equation in for the electric field, and the second term will be refereed to as the advection term. 
	
	Equation \eqref{OrigVlasov} can be exactly solved using the $G$-transform, which relies on the Hilbert transform.  So we review it and some of its properties next. 

	\subsection{Hilbert Transform Properties}
	\label{sec:Hilbert}
	
	The Hilbert Transform is defined by
	\bq
		H[g](v) \define \frac{1}{\pi} \dashint_\R \frac{g(u)}{u-v} \,du ,
	\label{HilbertDef} \eq
	where $\dashint$ is the Cauchy principle part of the integral.
	Following is a list of  some basic Hilbert transform properties, with well-known proofs (see e.g.\ Ref.~\onlinecite{King09}):  
	
	\bqy
		&&\bullet\ \  
		H[\ldots] \ \mathrm{is\ a\ linear\ operator.} 
		\label{HLinear} \\
		&&\bullet\ \  
		\mathrm{If}\  g(v)\ \mathrm{has\ a\ Hilbert\ transform,\
		then\ there\ is\  a\ function}\nonumber\\
		&& \qquad G,\ \mathrm{analytic\
		in\ the\ upper\ half\  complex}\ v-\mathrm{plane,\ 
		that} \nonumber \\
		&&\qquad \mathrm{limits\ to\ the\ real}\ v\ \mathrm{axis\ with} 
		\nonumber\\
		&&\qquad g(v) = g_R(v) + i g_I(v), \\
		&&\qquad \mathrm{where\ the\ real\ and\ imaginary\ parts\ of}\
		G\ \mathrm{are\ related\ by} \nonumber\\
		&&\qquad g_R = H[g_I] \ , \hspace{3em} g_I = - H[g_R]\,.
		\label{HAnalytic}\\
		&&\bullet\ \ 
		\mathrm{The\ inverse\ of\ the\ Hilbert\ transform\ is\ negative\ itself,\ i.e.} \nonumber\\
		&&\qquad H[H[g]] = -g \,. \label{HInverse} \\
		&&\bullet\ \ 
		\mathrm{For\ two\ functions}\ g_1\ \mathrm{and}\ g_2,\
		\mathrm{the\ following\ convolution} \nonumber\\
		&&\qquad \mathrm{identity\ holds:} \nonumber\\
		&&\qquad 
		H[g_1 H[g_2] + g_2 H[g_1]] = H[g_1] H[g_2] - g_1 g_2\,.
		\label{HConvId} \\
		&&\bullet\ \ 
		\mathrm{The\ Hilbert\ transform\ commutes\ with\
		differentiation,\ i.e.} \nonumber\\
		&&\qquad H\left[\frac{\p g}{\p u}\right] = 
		\frac{\p}{\p v} H[g]\,.	\label{HDeriv} \\
		&&\bullet\ \ 
		\mathrm{The\ Hilbert\ transform\ has\ the\ following\
		multiplication} \nonumber\\ 
		&&\qquad \mathrm{by}\ u\ \mathrm{identity:} \nonumber\\
		&&\qquad H[ug] = v H[g] + \frac{1}{\pi} 
		\dashint_\R g(u) \ du \,. \label{Hu} \\
		&&\bullet\ \ 
		\mathrm{The\ adjoint\ of\ the\ Hilbert\ transform\ is\
		negative\ itself,\ i.e.} \nonumber\\
		&&\qquad \dashint_\R f \ H[g] \ dv = 
		- \dashint_\R g \ H[f] \ dv \,. \label{Hdagger} \\
		&&\bullet\ \
		\mathrm{The\ Hilbert\ transform\ reverses\ the\ 
		parity\ of\ a\ function,\ i.e.} \nonumber\\
		&&\qquad \mathbf{\Pi} \Big[ H[f] \Big] = 
		-\mathbf{\Pi} [ f ] \,. \label{HParity}
	\eqy
	
Of interest in plasma physics is that fact that the  Hilbert transform of a Gaussian is the real part of the 
	plasma Z function, i.e. 
	\bqy
		Z\left(\frac{u}{v_t}\right) &:=& \pi
		H\left[\frac{1}{\sqrt{\pi}} e^{-v^2/v_t^2} \right]
		\ncr
		 &=& 
		\frac{1}{\sqrt{\pi}} \dashint_\R e^{-v^2/v_t^2} \frac{dv}{v-u} 
		\,.
		\label{ZDef}
	\eqy
	From Hilbert transform properties \eqref{HDeriv} and \eqref{Hu} 
	and from our knowledge 
	of the derivative of a Gaussian, we can determine that the
	derivative of the plasma Z function satisfies:
	\bqy
		\frac{\p Z}{\p u}  
		&=& - \frac{2}{v_t} \left( 1 + \frac{u}{v_t}
		Z\left(\frac{u}{v_t}\right) \right)
		\,.
		\label{ZDeriv}
	\eqy

	\subsection{$G$-Transform and Exact Linear Solution}
	\label{sec:LinSol}
	
	The $G$-transform is defined by 
	\bq
		f(v) = G[g](v) \define \ep_R(v) g(v) + \ep_I(v) H[g](v)\,, 
		\label{GDefinition}
	\eq
	where 
	\bq
		\ep_I(v) \define -\pi \frac{\om_P^2}{k^2} \frac{\p f_0(v)}{\p v}
		\ \mbox{and} \ 
		\ep_R(v) \define 1 + H[\ep_I](v) .
		\label{EpsDefinition}
	\eq
	In order for an integral transform to be useful, it has to have an
	inverse, which  for this one is given by
	\bq
		g(u) = G^{-1}[f](u) \define \frac{\ep_R(u)}{|\ep(u)|^2} f(u) - \frac{\ep_I(u)}{|\ep(u)|^2} H[f](u)
			\label{GInverse}\,, 
	\eq	
	where $|\ep|^2 \define \ep_R^2 + \ep_I^2$. 
	It is straightforward to check that $g(u) = G^{-1}[G[g]](u)$ 
	using the definition of $\ep_R$ \refeq{EpsDefinition} and 
	Hilbert transform properties \refeq{HInverse} and \refeq{HConvId}.
	
	The $G$-transform provides an exact solution to the one-dimensional 
	linear Vlasov equation. \cite{MorrPfir92,pjm94,Morr00}
	The transform takes you to new coordinates that naturally unravel 
	the phase mixing. In these new coordinates, there is
	no term from the electric field, so the equation is trivial.
	
	Apply the inverse $G$-transform in velocity space to \eqref{OrigVlasov} gives 
	\bqy
		&& \ppt G^{-1}[f_k] + i k \,G^{-1} [v f_k] 
		 \ncr
		&&\hspace{2cm}
		- i \frac{\om_P^2}{k} \int_\R f_k dv' \,G^{-1} \left[
		\frac{d f_0}{d v} \right] = 0\,.
		 \label{$G$-1Vlasov1}
	\eqy
	Defining  $g_k := G^{-1}[f_k]$, the first term is simple, 
	while the other two terms will take a little bit of work.	
	
	The second term can be dealt with using Hilbert transform property
	\refeq{Hu}. The correction term is the integral of a nonsingular 
	quantity, so we can remove  the principal value from  the integral, 
	\bqy
		G^{-1} [v f_k(v)] &=& \frac{\ep_R}{|\ep|^2} u f_k(u) - 
			\frac{\ep_I}{|\ep|^2} H[v f_k(v)] \ncr
		&=& \frac{\ep_R}{|\ep|^2} u f_k(u)
		\ncr
		&&\quad -
			\frac{\ep_I}{|\ep|^2} \left(u H[f_k(v)]
			+ \frac{1}{\pi} \dashint_\R f_k(v) dv \right) \ncr
		&=& u \left(\frac{\ep_R}{|\ep|^2} f_k(u) - 
			\frac{\ep_I}{|\ep|^2} H[f_k(v)] \right) 
			\ncr
			&&\quad  - \frac{1}{\pi} \frac{\ep_I}{|\ep|^2}
			\int_\R f_k(v) dv \ncr 
		&=& u \,G^{-1} [f_k] - \frac{\ep_I}{\pi |\ep|^2}
			\int_\R f_k(v) dv\,.
	\eqy
	The third term can be dealt with by recognizing that 
	${d f_0}/{d v}$ is only a constant away from $\ep_I$
	and that $H[\ep_I] = \ep_R - 1$.  Thus, 
	\bqy
		G^{-1}\left[\frac{d f_0}{d v}\right] &=&
			G^{-1}\left[-\frac{k^2}{\pi \om_P^2} \,\ep_I\right] = 
			-\frac{k^2}{\pi \om_P^2} G^{-1}[\ep_I] \ncr 
		&=& -\frac{k^2}{\pi \om_P^2} \left( \frac{\ep_R}{|\ep|^2}
			\ep_I - \frac{\ep_I}{|\ep|^2} (\ep_R - 1) \right)
			\ncr 
		&=& -\frac{k^2}{\pi \om_P^2} \frac{\ep_I}{|\ep|^2} \,.
	\eqy		
	
	Upon plugging these results back into \refeq{$G$-1Vlasov1}  we observe the 
	remarkable  cancellation
	\bqy
 \frac{\p g_k}{\p t}&=&  - i k \left(u g_k -
		\frac{\ep_I}{\pi |\ep|^2} \int_\R f_k dv \right)
		\ncr
		&&\qquad   
		+ \ i \frac{\om_P^2}{k} \int_\R f_k dv' \left(
		-\frac{k^2}{\pi \om_P^2} \frac{\ep_I}{|\ep|^2}\right) \ncr 
		&=&- i k u g_k \,.
		\label{Gadvec}
	\eqy
	The solution of \eqref{Gadvec} is trivial, 
	\bq
		g_k(u,t) = \initcond{g_k}(u) \,e^{-i k u t} \,.
		\label{$G$-1VlasovSol}
	\eq
	
	The one-dimensional linear Vlasov equation can be solved by 
	first inverse $G$-transforming the initial conditions to get
	$\initcond{g_k}(u)$,   using the solution \refeq{$G$-1VlasovSol}, and 
	then $G$-transforming back into the original coordinates, 
	\bq \label{VlasovSol}
		f_k(v,t) = G \left[ G^{-1}[\initcond{f_k}(v)] \,e^{-ikut} \right]
		\,. 
	\eq	
	It can be shown that this solution is equivalent to  van Kampen's solution, 
	which in turn is  equivalent to Landau's.
	
	\subsection{Landau Damping}
	\label{sec:Landamp}
	
	Evidently, the solution of \eqref{VlasovSol} must include Landau damping. If the equilibrium
	distribution function is monotonically decreasing, then the 
	spatial dependence of a perturbation (associated with the 
	electric field) decays. How does Landau damping appear in this
	solution?
	
	The distribution function itself doesn't damp - it could not 
	because this is a Hamiltonian system. Instead, what damps are  the
	density and electrical field perturbations, 
	which are proportional to the integral of the distribution function.  For example, 
	\bqy
		n_k(t) &=& \int_\R 
		G \left[ G^{-1}[\initcond{f_k}](u) \,e^{-ikut} \right] dv
		\nonumber\\
 &=& \int_\R \left( \ep_R(v) \,G^{-1}[\initcond{f_k}](v) \
			e^{-ikvt} \right.
			\ncr
			&+&  \left.
			  \ep_I(v) \,\frac{1}{\pi} \,\dashint_\R 
			G^{-1}[\initcond{f_k}](u) \,e^{-ikut} \frac{du}{u-v} \right)
			dv \ncr
		&=& \int_\R dv \,\ep_R(v) \,G^{-1}[\initcond{f_k}](v) \
			e^{-ikvt}
			 \label{RLMess}\\
			&+& \frac{1}{\pi} \,\dashint_\R du \int_\R dv \
			\frac{1}{u-v} \ep_I(v) \,G^{-1}[\initcond{f_k}](u) \
			e^{-ikut}\,.
			\nonumber
	\eqy
	Because the integrals in \eqref{RLMess}, including the integrals in the $G^{-1}$-transform,
	are not  simple, we will only look for the damping in the long time limit.
	
	The Riemann-Lebesgue lemma determines the long time limit; 
	for any sufficiently smooth function $F(\zeta)$, 
	\bq
		\lim\limits_{t \rightarrow \infty} \int_\R F(\zeta) \,
		e^{-i \zeta t} \,d\zeta = 0\,.
	\eq
	The integrals in \refeq{RLMess} are of the form specified in the 
	Riemann-Lebesgue Lemma, if we assume that the initial conditions 
	are sufficiently smooth and $|\ep|^2$ is never zero along 
	the real axis. We can thus conclude, 
	\bq
		\lim\limits_{t \rightarrow \infty} n_k(t) = 0\,.
	\eq
	The density perturbation of the plasma (and thus the electric field)
	decays to zero as time $\rightarrow \infty$. The Riemann-Lebesgue
	Lemma also tells us that the decay rate will be proportional to 
	the distance from the real axis to the nearest pole of $F(\zeta)$.
	In our case, that is the distance from the real axis to the nearest
	place where $|\ep|^2 = 0$.  From this one can obtain the usual damping rate.\cite{MorrPfir92}

\section{Collisions}
\label{sec:Collisions}

%

Let us now consider collisions together with the Vlasov-Poisson dynamics by adding a collision operator to the right-hand side of 
 \eqref{OrigVlasov}, 
		\bq 
		\label{VlasovCollisions}
		\frac{\p f_k}{\p t} + i k v f_k - i \frac{\om_P^2}{k}
		\int_\mathbb{R} f_k dv' \frac{d f_0}{d v}	= \calc[f]\,.
	\eq
Equation \eqref{VlasovCollisions} is no longer Hamiltonian.  A manifestation of this is that any good  the collision operator $\calc[f]$, will have  asymptotic stability to a Maxwellian  distribution.  A common choice for describing collisions is the 
	Fokker-Planck operator,\cite{Ray91,Fok14,Chan43,LenBern58,DegDes92} that  has the form
	\bq
		\calc[f] := 
		\nu \left(\frac{v_t^2}{2} \frac{\p^2 f_k}{\p v^2}
		+ v \frac{\p f_k}{\p v} + f_k \right) \,.
		\label{Coll}
	\eq
	This operator gives zero when it acts on the Maxwellian 
	\bq 
		f_0(v) = \frac{1}{\sqrt{\pi} v_t}e^{-v^2/v_t^2} . 
	\label{FPEquil} \eq
	If the advection and electric field terms are set to zero, then any other
	initial function of velocity will decay to this Maxwellian.
	
	Without collisions we had  a continuum of possible equilibria, but adding the 
	Fokker-Planck collision operator selects \refeq{FPEquil} as the
	only equilibrium.  Thus we no longer  work with general $\ep_I(v)$
	and $\ep_R(v)$ - instead we can be content with a special case  of the Maxwellian,  where 
	\bqy
		\ep_I 
		&=& \sqrt{\pi} \,\frac{2 \omega_P^2}{k^2 v_t^2} \,\frac{v}{v_t} 
		\,e^{-v^2/v_t^2}
		 \label{EpIMaxwell} \\
		\ep_R
		&=& 1 + \frac{2 \omega_P^2}{k^2 v_t^2} + \frac{2 \omega_P^2}
		{k^2 v_t^2} \,\frac{v}{v_t} \,Z\left(\frac{v}{v_t}\right) \,. 
		\label{EpRMaxwell}
	\eqy
	In \eqref{EpRMaxwell} $Z$ is the real part of the plasma Z function  of \refeq{ZDef}.
	
	If $\nu$ is small, we might be tempted to treat the collision
	operator as a perturbation on the original problem. 
	This is not easy because we are faced with a singular perturbation. The small
	parameter multiplies the highest derivative of $f$ with respect
	to $v$. If $f$ has structure on extremely small velocity scales, 
	then the highest derivative of $f$ can become $O(1/\nu)$, making 
	a conventional perturbation theory illegitimate. For the Vlasov-Poisson system, we are 
	guaranteed that $f$ will eventually get fine structure in $v$, since it behaves as 
	 $e^{-i k u t}$ for large $t$.
	
	This motivates our use of the  $G$-transform to attack this problem.	
	
	\subsection{$G$-Transform and Collisions}
	\label{sec:GColl}
	
	Now, let us apply the inverse $G$-transform in velocity space to
	\refeq{VlasovCollisions}.  Because we have already seen the cancellation 
	that occurs with the Vlasov part of this equation, we need only 
	examine its affect on  the collision operator,  which leads to  
	\bq
  \frac{\p g_k}{\p t} + i k u g_k = G^{-1} \left[ \calc [G[g_k]] \right]\,.
		\label{InitialGColl}
	\eq
	We will have to determine the commutation relations between the 
	$G$-transform and the collision operator. Before doing so we 
	state some properties of the collision operator. The proofs
	are all simple and some  indications of how to approach them are given 
	immediately after each.
\bqy
 &&\bullet  \   \  \calc[\ldots]\  \mathrm{is\  a\  linear\  operator.} 
  \label{CollLinear}\\
&&\bullet\  \  \calc[\ep_I] = -\nu \ep_I. 
 \label{CollEpsI} \\
		&&\qquad \mathrm{Recall}\  \ep_I \propto {d f_0}/{d v} \propto v \ \exp({-v^2/v_t^2}) \  \mathrm{and\  consider}
	\nonumber\\
		&&\qquad  \mathrm{the\  derivatives\  of\  this.}
\nonumber\\
&&\bullet\  \  \calc[AB] = B \, \calc[A] + A \, \calc[B] - \nu A B 
 			+ \nu v_t^2 
 		\frac{\p A}{\p v} \frac{\p B}{\p v}.
		 \label{CollProduct}
		 \\
		&&\qquad  \mathrm{Use\  the\  product\ rule\  several\  times\  and\  rearrange.}
		\nonumber \\
&&\bullet\  \ \calc[H[A]] = H[\calc[A]].
 \label{CollHilbert}\\ 
		&&\qquad \mathrm{Use\  Hilbert\  transform\  properties\  \eqref{HDeriv}\ and\  \eqref{Hu},\  and}
		\nonumber\\
		&&\qquad \mathrm{note\   everything\  decays\  at}\  \infty.
		\nonumber\\
%
&&\bullet \   \ \calc[\ep_R] = \nu (2 - \ep_R). \label{CollEpsR} \\
		&&\qquad \mathrm{Use\ the\  definition\  of}\  \ep_R\  \mathrm{of \ \eqref{EpsDefinition}\  and\  
		collision}
		\nonumber\\
		&&\qquad \mathrm{ operator\ property\ \eqref{CollEpsI}.}
		\nonumber
\eqy
	
	In order to evaluate $\calc[G[g_k]]$, we
	first use the linearity property \eqref{CollLinear}, 
	\bq
		\calc[G[g_k]] = \calc[\ep_R g_k] + 
		\calc[\ep_R H[g_k]] \,.
	\eq
Next  we use the collision operator product rule  \eqref{CollProduct} 
	on each term, and then apply the collision operator on $\ep_I$ using property \eqref{CollEpsI} and on $\ep_R$ using 
	property \eqref{CollEpsR},  
\bqy
		\calc[\ep_R g_k] &=& \ep_R \calc[g_k] + g_k \calc[\ep_R] - \nu \ep_R g_k + 
			\nu v_t^2 \frac{\p \ep_R}{\p v} \frac{\p g_k}{\p v} \ncr
		&=& \ep_R \calc[g_k] + 2 \nu (1 - \ep_R) g_k + 
			\nu v_t^2 \frac{\p \ep_R}{\p v} \frac{\p g_k}{\p v}\,,
\ncr
\calc[\ep_I H[g_k]] &=& \ep_I \calc[H[g_k]] + H[g_k] \calc[\ep_I] - 
			\nu \ep_I H[g_k] 
			\ncr
			&& \hspace{2cm} + 
			\nu v_t^2 \frac{\p \ep_I}{\p v} \frac{\p H[g_k]}{\p v} \ncr
		&=& \ep_I H[\calc[g_k]] - 2 \nu \ep_I H[g_k] 
		\ncr
		&&\hspace{2cm} + \nu v_t^2
		\frac{\p \ep_I}{\p v} H\left[\frac{\p g_k}{\p u}\right]\,.
\eqy
We can also commute the collision operator
	with the Hilbert transform using property \eqref{CollHilbert}, and note that a derivative can move inside of a Hilbert 
	transform using Hilbert transform property \eqref{HDeriv}.  Thus we obtain 
	\bqy
		\calc[G[g_k]] &=& \ep_R \calc[g_k] + 2 \nu (1 - \ep_R) g_k + 
		\nu v_t^2 \frac{\p \ep_R}{\p v} \frac{\p g_k}{\p v}\ncr
		&& + 
		\ep_I H[\calc[g_k]] - 2 \nu \ep_I H[g_k] + \nu v_t^2
		\frac{\p \ep_I}{\p v} H\left[\frac{\p g_k}{\p u}\right] \ncr
		&=& \big(\ep_R \calc[g_k] + \ep_I H[\calc[g_k]]\big) + 2 \nu g_k
		\ncr
		&& \hspace{1.6cm} - 2 \nu (\ep_R g_k + \ep_I H[g_k]) 
		\ncr
		&& \hspace{1.6cm}+ \nu v_t^2 \left(\frac{\p \ep_R}{\p v} 
		\frac{\p g_k}{\p v} + \frac{\p \ep_I}{\p v}
		H\left[\frac{\p g_k}{\p u}\right] \right) \ncr
		&=& G[\calc[g_k]] + 2\nu (g_k - G[g_k])\\
		&& \hspace{1.6cm}+ \nu v_t^2 \left(\frac{\p \ep_R}{\p v} 
		\frac{\p g_k}{\p v} + \frac{\p \ep_I}{\p v}
		H\left[\frac{\p g_k}{\p u}\right] \right) \,.
		\nonumber
	\eqy
	Using this result in \refeq{InitialGColl}, we can determine how the
	collision operator interacts with the $G$-transform, 
	\bqy
		\hspace{-.1cm}\frac{\p g_k}{\p t} + i k u g_k &=& G^{-1} \left[ \calc [G[g_k]] \right] \ncr
		&=& G^{-1} \bigg[ G[\calc[g_k]] + 2\nu (g_k - G[g_k]) \label{GCG} \\
		&& \hspace*{-1cm} +
			\nu v_t^2 \left(\frac{\p \ep_R}{\p v} 
			\frac{\p g_k}{\p v} + \frac{\p \ep_I}{\p v}
			H\left[\frac{\p g_k}{\p u}\right] \right) \bigg] \ncr
		&=& \calc[g_k] + 2 \nu G^{-1}[g_k] - \nu g_k 
		\\
		&& \hspace*{-1cm}  + 2 \nu v_t^2 \,G^{-1} \bigg[ \left(
		\frac{\p \ep_R}{\p v} \frac{\p g_k}{\p v} +
		\frac{\p \ep_I}{\p v} H\left[\frac{\p g_k}{\p u}
		\right]\right) \bigg]\nonumber\,.
	\eqy
	When dealing with the $G^{-1}$ of the last term, we use 
	Hilbert transform property \refeq{HConvId} to simplify the
	$H[...H[...]]$ term. 

Finally, we obtained the $G$-transformed one-dimensional 
Vlasov equation with collisions:
	\bqy
		\frac{\p g_k}{\p t} + i k u g_k &=& \calc[g_k] + 
		2 \nu (G^{-1}[g_k] - g_k) 
		\ncr
		&+&  \frac{\nu v_t^2}{|\ep|^2} 
		\bigg[ \left( \ep_R \frac{\p \ep_R}
		{\p u} - \ep_I \frac{\p \ep_I}{\p u} \right)
		\frac{\p g_k}{\p u} \ncr
		&+&  \left( \ep_R \frac{\p \ep_I}
		{\p u} + \ep_I \frac{\p \ep_R}{\p u} \right)
		H\left[\frac{\p g_k}{\p v}\right] \bigg]	\,.	
		\label{GTransformedVlasov}
	\eqy
	The left hand side of this equation is simply advection: 
	the electric field term	has vanished. The right-hand side of 
	this equation has the collision operator in terms of $u$, but it also has 
	other terms. The rest of the right-hand side describes how the electric field and the collisions interact. 
	We will call all of them the shielding term, $\cals[g_k]$.
	
	\subsection{Dropping the Shielding Term}
	\label{sec:DropShield}
	
	What have we gained by doing the $G^{-1}$-transformation?

	At first, it doesn't look like we've gained very much. 
	If the collision operator isn't there, the $G$-transform transforms 
	an integro-differential equation into a differential equation. 
	The equations of motion are dramatically simpler since they are local in $k$ and $v$.
	However, the shielding term that  arises from the 
	$G$-transform of the collision operator is also nonlocal. We have
	replaced an integro-differential equation with another
	integro-differential equation. And, the new one looks more complicated.
	
	However, if we know that some terms are small, 
	it is a reasonable approximation to drop all of the 
	complicated terms on the right-hand side and add them 
	in later as a perturbation.
	
	In most physical situations when the Vlasov equation is relevant,
	the collision frequency $\nu$ is assumed to be smaller than 
	the other frequencies in the system.
	\bq \nu g_k \ll \frac{\p g_k}{\p t} \ , \ k u g_k \eq	
	All of the terms on the right-hand side are proportional to $\nu$,
	so they all	appear to be smaller than the terms on the left hand
	side of \eqref{GTransformedVlasov}. This suggests that it might be possible to treat the 
	collision term as a perturbation in the original problem and 
	not worry about $G$-transforming the collision operator.
	
	Having a small collision frequency is not enough 
	to make sure that the right-hand side remains small. The collision 
	operator has a term proportional to ${\p^2 g_k}/{\p u^2}$, 
	which is the highest derivative with respect to $u$ in the problem.
	Although the left hand side may be originally dominant, the 
	dynamics create small-scale structure in velocity space, 
	with terms proportional to $\exp({i k v t})$. These small
	scales in velocity space make the collision operator significant,
	even for small collision frequencies.
	
	Note,  unlike the Fourier transform, the Hilbert transform
	preserves scale. If g(u) = H[f(v)](u) and $f$ is rapidly
	varying, say $f(v/\la)$ for $\la<<1$, then $g(u/\la)$.  
	Thus rapid variation is also preserved by the
	$G$-transform: a solution with a rapid scale of variation 
	$\exp({i k u t})$ in $u$ will have a  scale of variation 
	$\exp({i k v t})$ in $v$, and vice versa.

	The highest derivative in the shielding term is 
	${\p g_k}/{\p u}$. When a function has extremely small scales,
	its higher order derivatives are larger than its lower order
	derivatives. Even when the right-hand side of this equation is
	important, the collision operator still dominates the shielding term.
	
	This same argument could also be used for the last
	two terms of the collision operator \refeq{Coll}.
	They both are also multiplied by the small parameter $\nu$
	and neither has a second derivative. If we are 
	justified in dropping the shielding terms, then
	we are also justified in dropping the latter two 
	terms of the collision operator.
	
	We thus have two reasonable approximations that we could
	make: drop the shielding terms and keep the entire 
	collision operator or drop everything except the second
	derivative. We call the second derivative by itself
	$\cald[g_k]$ for diffusion and everything else can
	be lumped together into an expanded set of shielding
	terms $\cals'[g_k]$.
	
	If we keep only the second derivative, the resulting
	solution is easier to solve and analyze analytically.
	However, it's long-time limit is strange. The
	full collision operator relaxes to a Maxwellian with 
	width $v_t$. The collision operator $\cald$  results in a Gaussian
	whose width increases without bound. This is not too
	big of a concern for us since our perturbations 
	will decay on  a much shorter time than the collision time
	$1/\nu$. Numerically, these ever-expanding tails cause
	problems at the boundary. When we consider the problem 
	numerically in Sec.~\ref{sec:NumAdvDiff}, we use
	the entire collision operator. A comparison of the 
	analytic solutions for both of these reasonable approximations 
	in found in Appendix \ref{sec:app:CompColl}. 
	The two solutions agree anywhere they are significantly different from zero.
	
	The shielding term is a good term to initially ignore, and 
	then add back in as a perturbation.		
	
	\subsection{Exact Solution of the Advection-Diffusion Equation}
	\label{sec:ExactAdvDiff}

	Given the arguments of Sec.~\ref{sec:DropShield}, we proceed with the local equation:
	\bq
		\frac{\p g_k}{\p t} + i k u g_k = \nu \frac{v_t^2}{2} 
		\frac{\p^2 g_k}{\p u^2}\,,
		\label{AdvDiffEq}
	\eq
	an equation that can be exactly solved. First Fourier transform in velocity. 
	This reduces the equation to being first order in time and velocity instead of 
	being first order in time and second order in velocity. We can then use the method
	of characteristics to solve this problem with the Fourier transformed initial 
	conditions. Afterwards, we inverse Fourier transform back
	into the original velocity coordinate. 
	Instead of working through this calculation, we will just
	state the solution.
	For more details, see Sec.~\ref{sec:ShieldPert}, where we 
	consider the corresponding inhomogeneous problem.
	Given the equation \refeq{AdvDiffEq} and an initial
	condition which is Gaussian in velocity and has a wave number $k$ 
	in position, 
	\bq
		g_k(u,0) = \frac{1}{\sqrt{\pi} v_t} e^{ -{u^2}/{v_t^2}}
		\,, \label{GaussIC}
	\eq
	the solution to \eqref{AdvDiffEq} is
	\bqy
		g(u,t) &=& \frac{1}{\sqrt{\pi} v_t \sqrt{1+2\nu t}}
		\exp\bigg[-\frac{u^2}{v_t^2 \,(1+2\nu t)} \label{SolNoShield}\\
		 &&
		- i k u t \,\frac{1+\nu t}{1+2\nu t}
		- \frac{1}{12} k^2 v_t^2 \nu t^3 \,
		\frac{2+\nu t}{1+2\nu t} \bigg]\,.
		 \nonumber
	\eqy
	As a check, it is not difficult to plug this solution back into the equations and 
	see that it does satisfy them. An existence and uniqueness
	theorem guarantees that this is the only correct solution.
	
	The exponential of \eqref{SolNoShield} has three terms. The first term is the
	result of the diffusion part of the equation. If we
	ignore advection here, the remaining heat equation will
	cause a Gaussian to spread out in time. The variance in
	velocity space increases, but the amplitude of the 
	perturbation does not change. The time scale for this 
	process is $1/\nu$. In our ordering, the perturbation will
	already have decayed away before this increased variance 
	becomes significant.
	
	The second term is primarily the result of the advection 
	part of the equation. It is a velocity dependent phase 
	shift of the initial conditions. This doesn't change the
	amplitude of the perturbation to the distribution function.
	As we saw in Sec.~\ref{sec:Landamp}, it does result 
	in Landau damping for the perturbation to the density
	and electric field.
	
	The rate at which this occurs is modulated by an extra
	factor resulting from the collision operator. This 
	factor is unity at $t=0$ and decays to $1/2$ as 
	$t\rightarrow\infty$. 
	This change occurs with a time scale $1/\nu$, so the 
	perturbation will already have damped away before its 
	effect becomes significant. However, as we will see in 
	Sec.~\ref{sec:OIC}, this effect can occur much sooner
	for other initial conditions.
	
	The third term is the result of the interaction between
	the two terms. It is also where the damping of the 
	perturbation to the distribution function comes from. 
	Neither term individually damps the distribution function,
	but their interaction does.
	
	This term is modulated by a factor which varies from $2$ 
	at $t=0$ to $1/2$ as $t\rightarrow\infty$. The time
	scale over which this happens is $1/\nu$, so it will be
	close to $2$ until the perturbation has almost entirely
	decayed.
	
	If we define the decay time, $t_D$, for this equation to be the
	time for the perturbation to the distribution function 
	to decay to $1/e$ of its original value, and  then  approximate
	the modulating factor to be $2$, we find 
	\bq
		-1 = -\frac{1}{6} k^2 v_t^2 \nu t_D^3 \ \Rightarrow
		\ t_D = \left(\frac{6}{k^2 v_t^2 \nu}\right)^{1/3}\,.
		\label{DecayTime}
	\eq
	Since this goes as $\nu^{-1/3}$, this time is much shorter 
	than the time scale associated with the collisions by
	themselves.
	Callen also found an effective damping rate which scales 
	as the (parallel) collision frequency to the $1/3$ 
	when considering a similar problem.\cite{Call14}
		
	\subsection{Advection-Diffusion Crossover}
	\label{sec:AdvDiffCross}
	
	We can take our solution and plug it back into each of the terms of
	\refeq{GTransformedVlasov}. Doing so will tell us when each of 
	the terms dominates - when advection dominates the damping and 
	when collisions dominate the damping.
	
	It is easy to evaluate the terms in \refeq{AdvDiffEq} 
	on the solution:
	\bqy
		\cala[g_k] &=& i k u \,g_k \\
		\cald[g_k] 
		&=& \frac{\nu}{(1+2\nu t)^2} \Big(2 \frac{u^2}{v_t^2} - 
		\Half v_t^2 k^2 t^2 (1 + \nu t)^2 \ncr 
		& &\quad  - (1+2\nu t) + 2 i k u t (1+\nu t)\Big) \,g_k\,, 
	\eqy
	whence we can calculate the ratio of the diffusion term to the 
	advection term.  Nondimensionalizing the result using 
	$\varep = \nu / k v_t$ , $\tau = k v_t t$ , and 
	$\xi = u / v_t$ gives
	\bqy
		\frac{\cald[g_k]}{\cala[g_k]} &=& \frac{-i \,\varep}{\xi
		(1 + 2 \varep \tau)^2} \Big(2 \xi^2 - 
		\Half \tau^2 (1 + \varep \tau)^2 \ncr 
		& & - (1+2\varep\tau) + 2 i \xi \tau (1+\varep\tau)\Big)\,.
		\label{ratio}
	\eqy
	
\begin{figure}[htb]
		\includegraphics[scale=.35]{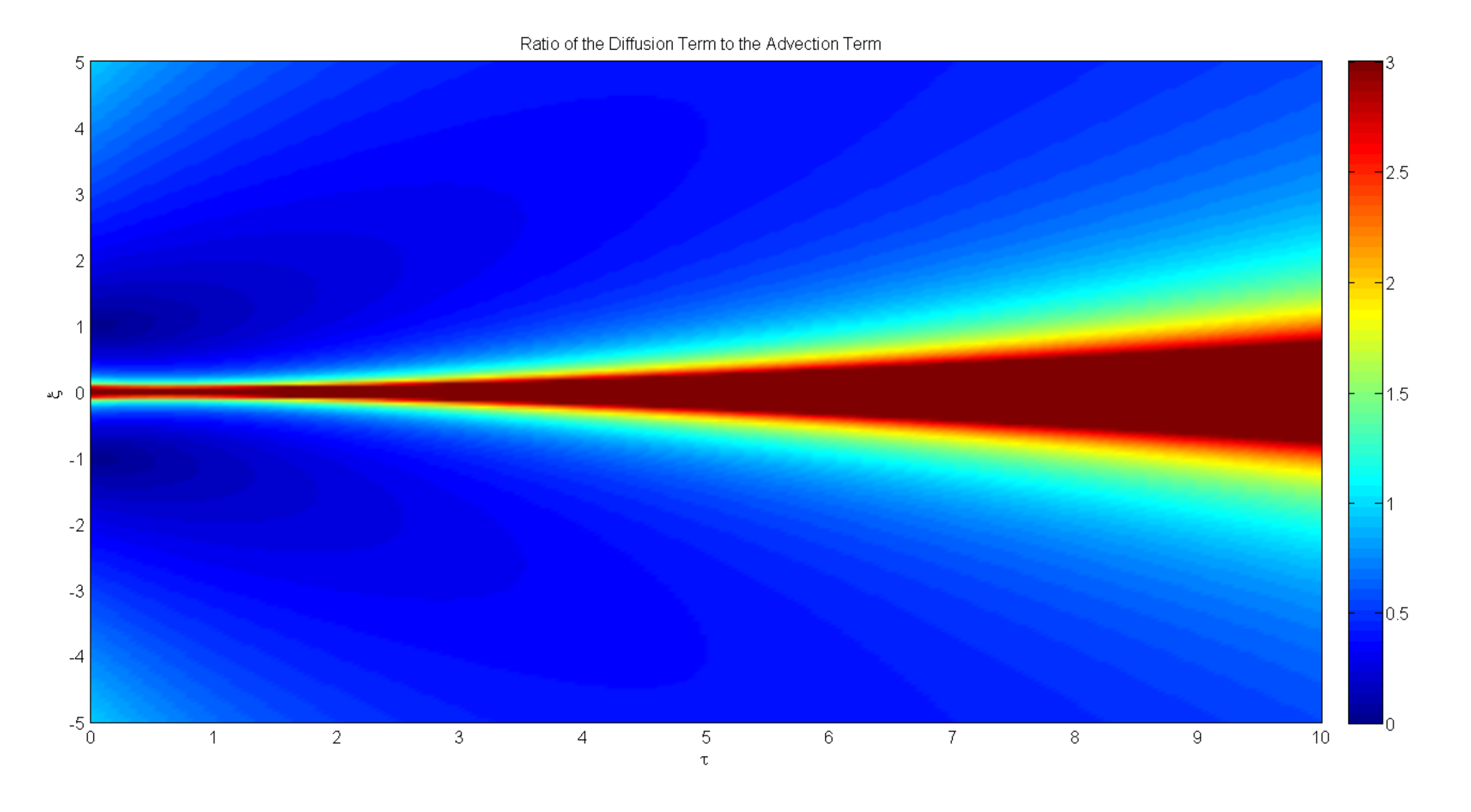}
		\caption{The magnitude of the ratio of the collision term to the advection term,
		plotted as a function of non-dimensional
		time and velocity, respectively.}
		\label{fig:Coll2Adv}
	\end{figure}

	In Fig.~\ref{fig:Coll2Adv}, we plot the magnitude of 
	the ratio of \eqref{ratio} for various nondimensionalized times and 
	velocities to see when each term dominates and 
	when the two terms are comparable. For this plot, we fix
	$\varep = 0.1$.
	
	The advection term dominates for small times and large
	velocities.	The collision term dominates for large times
	and small velocities. For sufficiently large velocities,
	the collisions start becoming important again, but
	the distribution function is so small at these velocities
	that this isn't significant.
	
	\subsection{Other Initial Conditions}
	\label{sec:OIC}
	
	The perturbation to the distribution function could 
	be a very different temperature than the equilibrium 
	temperature that appears in the collision operator.
	Let us choose as an initial condition in $u$ a Gaussian 
	with a different initial thermal velocity, say $v_I$, 
	\bq
		g_k(u,0) = \frac{1}{\sqrt{\pi} v_I}
		e^{ -{u^2}/{v_I^2}}\,.
		\label{ColdGaussIC}
	\eq
	The calculation proceeds as before. Upon defining $\alpha :=
	v_I^2 / v_t^2$, we find 
	\bqy
		g(u,t) &=& \frac{1}{\sqrt{\pi} v_t \sqrt{\al+2\nu t}}
		\exp\bigg[-\frac{u^2}{v_t^2 \,(\al+2\nu t)}  \label{ColdSolNoShield}\\
		 &&
		- i k u t \,\frac{\al+\nu t}{\al+2\nu t}
		- \frac{1}{12} v_t^2 k^2 \nu t^3 \,\frac{2\al+\nu t}
		{\al+2\nu t} \bigg]\,.
		 \nonumber
	\eqy	
	This has the same asymptotic behavior as before, but the 
	time it takes for the modulation factors 
	to transition from close to unity to close to 
	$1/2$ is now $\alpha/\nu$. This transition is
	observable if this time is shorter than the decay time, i.e. 
	\bq
		\frac{\alpha}{\nu} \lesssim t_D \ \Rightarrow \
		\frac{v_I}{v_t} \lesssim 
		\left(\frac{\nu}{k v_t}\right)^{1/3}\,.
	\eq
	In order to measure the modulation of the phase mixing 
	term, you have to use initial conditions with thermal
	velocities  much smaller than the equilibrium thermal
	velocity.
	
	We could also consider initial conditions in $u$ that
	are a polynomial times a Gaussian. The calculation 
	proceeds as before - the only difference is that there
	is now a polynomial multiplying everything. The 
	solution will also be \refeq{SolNoShield}, multiplied
	by some polynomial in $u$ and $t$. 
	
	An appropriately chosen set of polynomials, such as the
	Hermite polynomials, will form a complete basis. 
	Any other bounded function that
	decays quickly enough in velocity can be written as a
	superposition of these functions. Our analysis of
	the solution in Sec.~\ref{sec:ExactAdvDiff} also 
	applies for many initial conditions.

	\subsection{Realistic Initial Conditions}
	\label{sec:RIC}
	
	Gaussian initial conditions in velocity are not physically 
	realistic in $u$ space. The initial conditions
	\refeq{GaussIC} are commonly used in kinetics because they
	represent a purely spatial perturbation to the distribution
	function. We have transformed the velocity coordinate, so
	we should use different initial conditions for the
	advection-diffusion equation, viz. 
	\bq
		\initcond{g_k}(u) = G^{-1}\left[\initcond{f_k}(v)\right] = 
		G^{-1}\left[\frac{1}{\sqrt{\pi} v_t}
		e^{-{v^2}/{v_t^2}}\right] .
		\label{RealIC1}
	\eq
	
	We assume that the thermal velocity for the equilibrium is the same 
	as the thermal velocity of the initial conditions. We've already 
	calculated $\ep_I$ \refeq{EpIMaxwell} and $\ep_R$ \refeq{EpRMaxwell}.
	Thus the  initial condition of \eqref{RealIC1} in $G^{-1}$-transformed space is 
	\bqy
		\initcond{g_k}(u) 
		&=& \frac{\frac{1}{\sqrt{\pi} v_t}	e^{-u^2/v_t^2} 
		(1 + \frac{2 \omega_P^2}{k^2 v_t^2})} 
		{A^2 + B^2}\,,
		 \label{RealIC}
	\eqy
	where
	\bqy
	A&:=& \sqrt{\pi}
		\frac{2 \omega_P^2}{k^2 v_t^2} \frac{u}{v_t}
		e^{-u^2/v_t^2} 
	\ncr
	B&:=& 1 + \frac{2\omega_P^2}
		{k^2 v_t^2} +  \frac{2\omega_P^2} {k^2 v_t^2} \frac{u}{v_t}
		Z({u}/{v_t}) \,.
	\eqy
	We will use \eqref{RealIC} in Sec.~\ref{sec:NumAdvDiff}. 
	
	\subsection{Numerically Comparing Advection and Diffusion}
	\label{sec:NumAdvDiff}
	
	Instead of trying to deal with the initial conditions 
	\refeq{RealIC} analytically, we solved the advection-diffusion equation with the entire collision 
	operator \refeq{AdvDiffEntireColl} numerically 
	using a finite difference method.  Our code is written in MATLAB, and is simple enough to be run on a personal laptop.
	
	A  Gaussian equilibrium distribution function,
	\[
	{f_0}(v) = 
	\frac{1}{\sqrt{\pi} v_t} e^{-v^2/v_t^2}\,,
	\]
	with units in velocity and position where $v_t = 1$ and $k = 2\pi$ was chosen.
	In the code we used 200 cells in $x$ with periodic boundary conditions and 
	200 cells in $v$ that range between $\pm 2.5 v_t$  with Dirichlet boundary conditions.
	In our units of time, the collision frequency is 
	$\nu = 0.1$ and the plasma frequency is $\omega_P = 5$.
	Our time step was $0.01$ and we ran the advection-diffusion
	equation until $t = 2$, at which point the initial
	perturbation had almost completely decayed. The decay 
	time \refeq{DecayTime} for these parameters is $t_D = 1.15$.
	The non-dimensional parameters for this run are
	\bq
		\frac{k v_t}{\nu} = 20\pi \ , \ \frac{\omega_P}{\nu} = 50 .
	\eq
	
	To construct the realistic initial conditions \refeq{RealIC}, 
	we used MATLAB's built in Hilbert transform to numerically apply 
	the $G^{-1}$-transform to Gaussian initial conditions in the
	original coordinates. Any other initial conditions for the original
	coordinates could be similarly numerically $G^{-1}$	transformed.
	After finding the solution, we numerically apply the $G$-transform
	to get the solution in the original velocity coordinate.
	
	We emphasize that our finite difference solution of the advection-diffusion equation  
	did not include the shielding term since it should be
	small. This assumption is checked below.
	
	We can compare the sizes of various terms in both the
	original coordinates and in the $G^{-1}$-transformed
	coordinates. In both velocity coordinates, there is a time
	derivative, an advection term, and a Fokker-Planck
	collision operator. Even though the terms look the same,
	the solution is different in the two coordinates, so we
	will get different results when we evaluate the term on the
	solution. The original coordinates also have a term from
	the electric field and the transformed coordinates also
	have the shielding term.
	
	We take the solution for the advection-diffusion equation 
	and evaluate it on all of the terms 
	in both coordinates. The shielding term, evaluated on the solution
	calculated without it, should be small. We also $G$-transform the
	shielding term evaluated on the solution back into the original
	coordinates,  so we can see how significant what we neglected is there.
	
\begin{figure}[htb]
\centering
\subfigure[{\footnotesize \  }]{\includegraphics[width=0.4\textwidth]{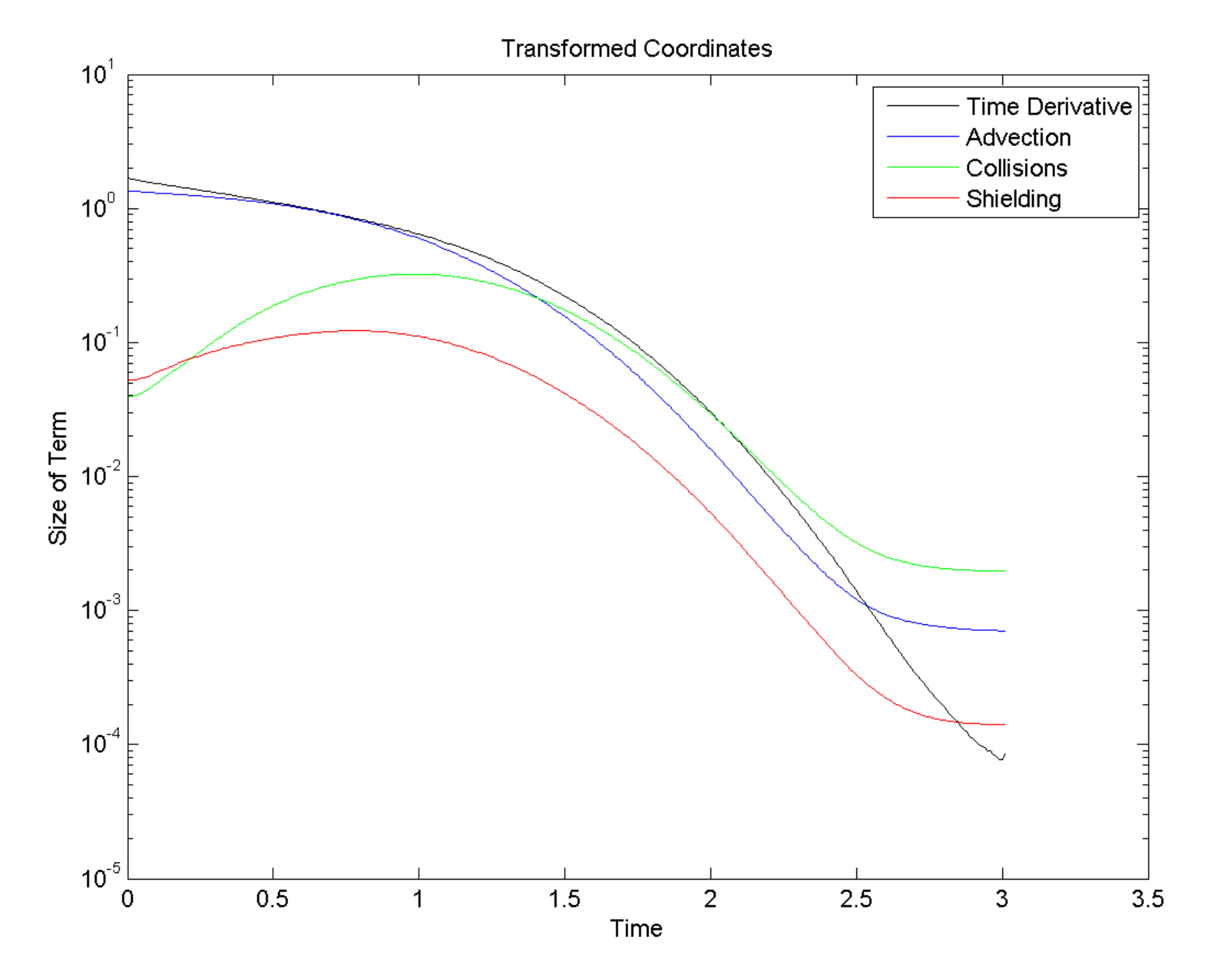}
\label{fig:CompTermsa}
}
\subfigure[{\footnotesize \  }]{\includegraphics[width=0.4\textwidth]{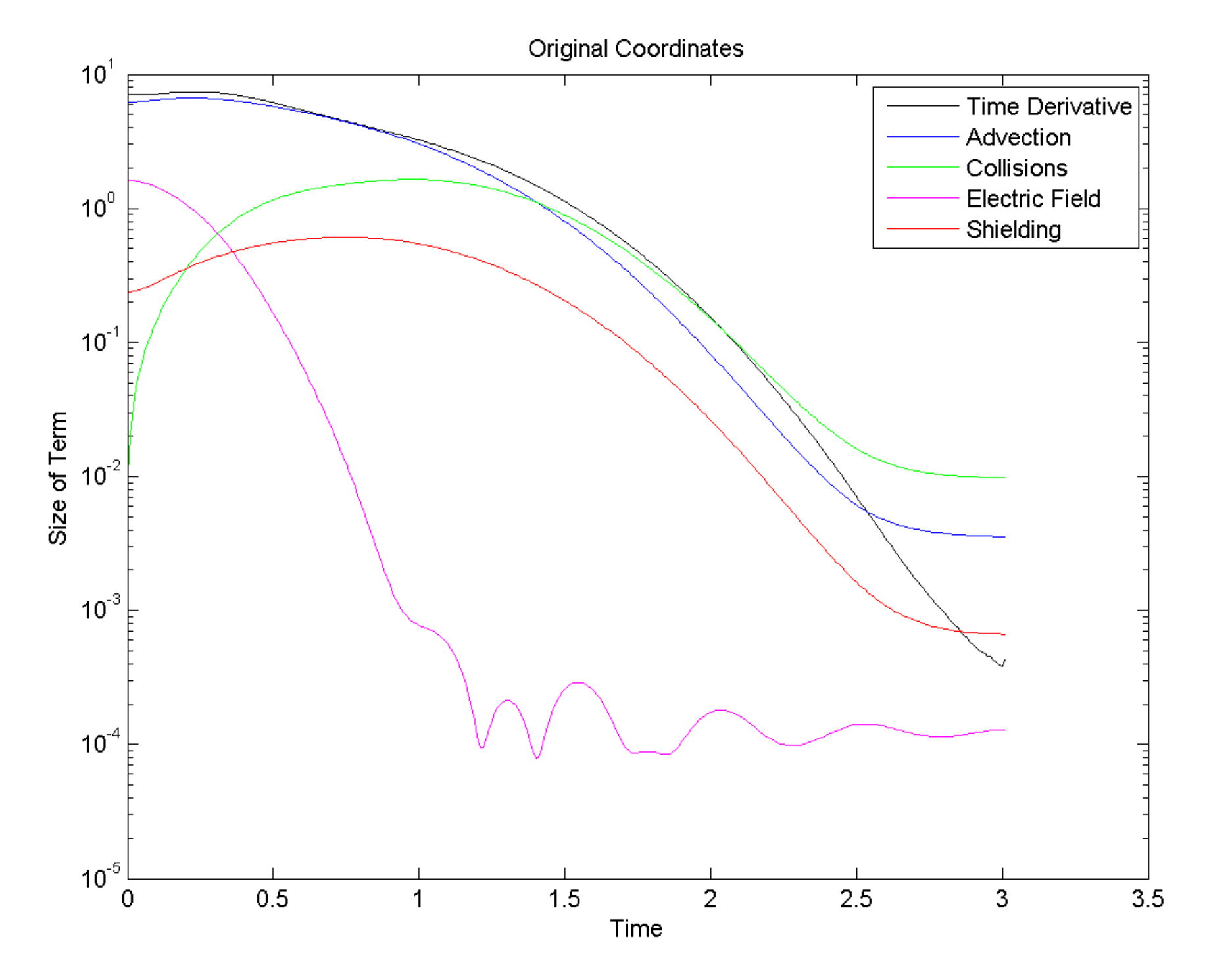}
\label{fig:CompTermsb}
}
\caption[sm]
{The time dependence of each of the terms of 
		the collisional Vlasov equation, $L^1$ 
		integrated in position
		and velocity. (a) In the $G^{-1}$-transformed coordinates, 
		there are four terms: a time derivative, an advection term, 
		a Fokker-Planck collision operator, and the shielding term. 
		Neglecting the shielding term when finding the solution is
		justified since the shielding term is always at least an order
		of magnitude smaller than the dominant term. (b) In the 
		original coordinates, there are four terms: a time derivative,
		an advection term, a Fokker-Planck collision operator, 
		and an electric field term. We also $G$-transformed the shielding
		term evaluated on the solution to see how significant the
		dropped term is. It is always at least an order of magnitude
		below the dominant term.}
\label{fig:CompTerms}
\end{figure}

	We $L^1$ integrate all of these terms in both position and velocity 
	and plot the resulting magnitudes of each as a function of time 
	in Fig.~\ref{fig:CompTerms}.

	For small times, the advection term is dominant in both the 
	original and transformed coordinates. In the original coordinates,
	the electric field term is the same order of magnitude as the
	advection term. One way that we could have reached the (local)
	advection-diffusion equation is by dropping the electric
	field terms and not $G$-transforming. We see here that this plan is
	illegitimate. Increasing the plasma frequency increases the
	significance of the electric field at small times.
	
	For large times, the collision term is dominant in both the 
	original and transformed coordinates. This is unsurprising since 
	the advection term creates small scale structures in velocity space.
	The collision operator has the highest order velocity derivative in
	the equation, so it becomes more significant when there is lots 
	of small scale structure.
	
	At an intermediary time, there is a crossover
	where collisions replace advection as the dominant term.
	
	The shielding term is always at least an order of magnitude lower
	than the other terms. This gives us confidence that neglecting it 
	is reasonable.
		
	\subsection{Introducing the Shielding Term as a Perturbation}
	\label{sec:ShieldPert}
	
	Since we know that the shielding terms and part
	of the collision operator will be small compared to 
	the other terms, we can stick a small parameter, $\de$, 
	in front of these terms and write the solution $g$ as 
	a series in this parameter, 
	\bq 
	g_k = g_k^{(0)} + \de g_k^{(1)} + \de^2 g_k^{(2)} + \cdots 
	\,.
	\eq
	The resulting hierarchy of equations reads
	\bqy
		\frac{\p g_k^{(0)}}{\p t} + i k u g_k^{(0)} &=& \cald[g_k^{(0)}] \\
		\frac{\p g_k^{(1)}}{\p t} + i k u g_k^{(1)} &=& \cald[g_k^{(1)}] 
			+ \cals'[g_k^{(0)}] \\
		\frac{\p g_k^{(2)}}{\p t} + i k u g_k^{(2)} &=& \cald[g_k^{(2)}]
			+ \cals'[g_k^{(1)}] \\
		&\vdots& \nonumber
	\eqy
	where $\cald$ and $\cals'$ were defined  in Sec.~\ref{sec:DropShield}.
	
	The zeroth order equation is the one solved above. 
	All of the equations of other orders are inhomogeneous
	versions of the one solved above. The only addition is a
	known function of velocity, determined by the lower order
	solutions, added to the right-hand side.	
	
	The initial conditions of the higher order terms are 
	all taken to be zero, i.e., the initial conditions are 
	entirely included in the zeroth order equation.
	
	Upon Fourier transforming in velocity space, $u \rightarrow \eta$, letting  
	$\hat{g} := \calf[g_k^{(n)}]$ and $\hat{S} :=
	\calf[\cals'[g_k^{(n-1)}]]$ gives 
	\bq
		\frac{\p \hat{g}}{\p t} + k \frac{\p \hat{g}}{\p \eta}
		+ \Half \nu v_t^2 \eta^2 \hat{g} = \hat{S}(\eta)\,.
		\label{xform}
	\eq
	We solve \eqref{xform} using  the method of characteristics, leading to  the following set of
	differential equations for $t(r,s), \eta(r,s),$ and $\hat{g}(r,s)$:
	\bqy
		\frac{dt}{ds} &=& 1\,,   \qquad \frac{d\et}{ds} = k ,
		\\
		\frac{d\hat{g}}{ds} &=& -\Half \nu v_t^2 \eta^2 \hat{g}
		+ \hat{S}(\eta) \\
		t(r,0) &=& 0 \,,   \qquad \eta(r,0) = r \, ,   \quad \hat{g}(r,0) = 0
	\eqy
	These equations are straightforward to solve. 
	The first two are trivial, 
	\bq 
		t(r,s) = s , \hspace{2em} 
		\eta(r,s) = k s + r \,.
		\label{TandEta}
	\eq
	Next we plug these into the equation for $\hat{g}(r,s)$, 
	and solve it using an integrating factor. Since we
	explicitly know what $\eta(r,s)$ is, we can evaluate 
	the integrals in the exponentials. We cannot evaluate the
	integral $ds''$ unless we specify explicitly what
	$\calf[\cals'[g_k^{(0)}]]$ is.	  Thus, 
	\bqy
		\frac{d}{ds} \left[\hat{g} \,e^{-\nu \frac{v_t^2}{2}
		\int\limits_0^s \eta^2(s') ds'} \right] &=& 
		e^{-\nu \frac{v_t^2}{2} \int\limits_0^s \eta^2(s'))
		ds'} \,\hat{S}(\eta(s)) \nonumber
	\eqy
	\bqy
		\hat{g}(r,s) &=&  e^{\nu \frac{v_t^2}{2}
		\int\limits_0^s \eta^2(s') ds'}
		\int\limits_0^s e^{-\nu \frac{v_t^2}{2}
		\int\limits_0^{s''} \eta^2(s''') ds'''}
		\hat{S}(\eta(s'')) \,ds'' \ncr 
		&=& \exp\Big[\frac{\nu v_t^2}{6 k}(k^3 s^3 + 3 k^2 s^2 r
		+ 3 k s r^2)\Big] \ncr 
		& & \times \int\limits_0^s 
		\exp\Big[-\frac{\nu v_t^2}{6 k}(k^3 s''^3 + 
		3 k^2 s''^2 r + 3 k s'' r^2)\Big] \ncr
		& & \qquad \times \,\hat{S}(k s'' + r) \,ds''\,.
	\eqy
	
    We now invert \refeq{TandEta} to get $s(t,\eta)$ 
    and $r(t,\eta)$, 
    \bq
    	s = t , \hspace{2em} r = \eta - k t\,,
    \eq
     and  substitute these in to get $\hat{g}(\eta,t)$. 
	Note that we do not substitute anything in for $s''$ since it is the 
	integration variable. From this point onward, we refer to it as $s$.  With the above we obtain     
	\bqy
		\hat{g}(\eta,t) &=& \exp\Big[
		\frac{\nu v_t^2}{6 k}(3 \eta^2 k t - 
		3 \eta k^2 t^2 + k^3 t^3)\Big] \ncr 
		& & \times \int\limits_0^t 
		\exp\Big[-\frac{\nu v_t^2}{6 k}(3 \eta^2 k s - 
		3 \eta k^2 s^2 + k^3 s^3)\Big] \ncr
		& & \qquad \times \,\hat{S}(\eta + k(s-t)) \,ds\,.
		\label{solgeta}
	\eqy
	Finally the  inverse Fourier transform of \eqref{solgeta}  in velocity space yields $g_k^{(1)}(v,t)$.
	We will not write this out explicitly since our expressions are long enough already.
	
	The higher order corrections can be done using exactly the 
	same technique. 
	
	\subsection{Other Collision Operators}
	\label{sec:OtherColl}

	The strategy for any collision operator is the same. 
	We show that the commutator between the collision operator and the 
	$G$-transform is small, then neglect it. The dramatic simplification
	of the left hand side only creates a small correction of the 
	right-hand side.
	
	If we choose some local collision operator with a finite number of
	velocity derivatives, say 
	\bqy
		Q_{n}[g_k] := c \frac{\p^n}{\p v^n} g_k +
		\calo\left[\frac{\p}{\p v}\right]^{n-1} \hspace{-.3cm}g_k\,,
	\eqy
	then we may consider how this collision operator acts on the $G$-transform.
	\bqy
		Q_n[G[g_k]]
		&=& c \frac{\p^n}{\p v^n} \Big(\epsilon_R  g_k + 
			\epsilon_I H[g_k]\Big) + 
			\calo\left[\frac{\p}{\p v}\right]^{n-1} \hspace{-.25cm}g_k \ncr
		&=& \epsilon_R c \frac{\p^n g_k}{\p v^n} + \epsilon_I
			H\left[c \frac{\p^n g_k}{\p v^n}\right] +
			\calo\left[\frac{\p}{\p v}\right]^{n-1} \hspace{-.25cm}g_k \ncr
		&=& G\left[c \frac{\p^n g_k}{\p v^n}\right] +
			\calo\left[\frac{\p}{\p v}\right]^{n-1} \hspace{-.25cm}g_k \ncr
		&=& G[Q_n[g_k]] + \calo\left[\frac{\p}{\p v}\right]^{n-1} \hspace{-.25cm} g_k
	\eqy
	The commutator between $Q_n$ and $G$ is 
	$\calo\left[{\p}/{\p v}\right]^{n-1} g_k$.
	As long as the original collision operator is multiplied by a
	small collision frequency, this shielding term will always be 
	much smaller than either the advection term or the collision term.
	It can be neglected. The only problems could arise when $g_k$ 
	is close to the kernal of $Q_n$ and the velocity is small.

\section{Gyrokinetics and Drift-Kinetics}
\label{sec:Gyrokinetics}

	A three-dimensional highly magnetized plasma is often 
	approximated by gyrokinetic or drift-kinetic equations. 
	\cite{FriChen82,BrizHa07,Krom12}
	We consider small electrostatic 
	perturbations around an equilibrium with slab geometry - the
	equilibrium has no electric field and the magnetic field is 
	constant and pointed in the $z$ direction. The motion of the ions
	is driven by an equilibrium density and temperature gradient in the 
	$x$ direction, while the motion of the electrons is determined by 
	quasineutrality. Since most of the particles' 
	motion is gyration about magnetic field, we can integrate over
	the gyrophase and use the adiabatic invariance of the magnetic
	moment to reduce the number of velocity
	dimensions to one: the velocity along the magnetic field lines.
	To make this approximation, we have to assume that the equilibrium 
	fields do not vary on a length scale shorter than the 
	Larmor radius and that the relevant time scales are long compared
	to the Larmor frequency. The perturbed quantities are allowed
	to vary on a length scale comparable to the Larmor radius, so the
	$\vec{E} \times \vec{B}$ nonlinearity is significant.

	The system obtained by the above approximations is the subject of  ongoing numerical studies.
	\cite{HatJen14,SchDor16} Specifically, the equations, in three spatial dimensions, one velocity
	dimension, and time, are
			\bq
			\label{Gyro}
 \frac{\p f}{\p t} + v \frac{\p f}{\p z} + 
		\frac{\p \varphi}{\p z} \,v F_M + \Half \rho_i v_t \
		[\varphi, f]_{x,y}  =   \calc[f] + \chi  \,, 
\eq
%
	where $\rho_i = {m v_{t}}/({e B})$, 
	\bq
		\varphi = \frac{Z T_e}{T_i} \int f \,dv \,, \quad  
		F_M = \frac{1}{\sqrt{\pi} v_t} e^{-v^2 / v_t^2} \,, 
	\eq
and 
	\bq \label{GyroSource}
		\chi = \left(-\frac{\rho_i v_t}{2} \frac{\p \varphi}{\p y}
		\right) \!\left(\frac{1}{L_n} + \left(\frac{v^2}{v_t^2} -
		\frac{1}{2}\right) \frac{1}{L_T}\right) F_M,
	\eq
	with $L_n, L_T$ being length scales for density and temperature
	gradients, respectively.  The nonlinear Poisson bracket
	term of \eqref{Gyro} is  the perpendicular advective	derivative, i.e.,
\bqy
		\vec{V}_\perp \cdot \nabla_\perp f &=& 
		\Half \rho_i v_{t} \left( \hat{z} \times \nabla_\perp
		\varphi \right) \cdot \nabla_\perp f \ncr
		&=&	\Half \rho_i v_{t} \,[\varphi, f]_{x,y}\,.
	\eqy

	\subsection{$G$-Transforming Gyro-/drift-kinetics}
	\label{sec:ModG}
	
	The gyrokinetic and drift-kinetic equations can be
	simplified using a slightly
	simple form of the $G^{-1}$-transform. In drift-kinetics,
	quasineutrality replaces the Poisson
	equation, so there are no derivatives in the relationship 
	between $\varphi$ and $\int f \,dv$. 
	We do not have to Fourier transform this equation in
	position before applying the $G^{-1}$
	transform; the $\ep's$ are independent of $k$.
	
	The only change is in the definition of $\ep_I$:
	\bq
	\ep_I(v) \define \pi \,\frac{Z T_e}{T_i} \,v F_M(v) \,.
	\label{EpsDefinitionMod}
	\eq
	The rest of the $G$-transform and its inverse for
	gyrokinetics are the same as before (\ref{GDefinition}-\ref{GInverse}).
	
	The gyrokinetic equations are extremely similar. 
	Instead of evaluating the fields at the gyrocenter,
	the fields are evaluated at a gyroradius away from the 
	gyrocenter. This introduces some additional dependence
	on the perpendicular spatial directions to the Poisson
	equation. $\epsilon_I$ will reflect this as well:
	\bq
		\epsilon_I(v) = \frac{\pi \, v F_M(v) \, e^{-k_\perp^2}}
		{\frac{T_i}{Z T_e} + 1 - 
		e^{-k_\perp^2} I_0(k_\perp)} \,,
		\label{EpsDefGyro}
	\eq	
	where $I_0$ is the zeroth order modified Bessel function.
	This does depend on $k_\perp$, but it does not require 
	you to Fourier transform in the parallel spatial direction.
	
	We will focus on the drift-kinetic equations because they
	are simpler and the calculations proceed similarly.

	We take the $G^{-1}$-transform
	of all terms of  \eqref{Gyro} and realize that the spatial and time derivatives commute
	with the $G^{-1}$-transform, as does any function of only space
	and time, such as $\varphi$. 
%
We define $g := G^{-1}[f]$ and  consider each term of \eqref{Gyro}.  The first term is simple.  
The second term can be dealt with using Hilbert transform property \eqref{Hu}, 
	\bqy
		G^{-1} [v f] 
		&=& \frac{\ep_R}{|\ep|^2} u f - 
			\frac{\ep_I}{|\ep|^2} \left(u H[f]
			+ \frac{1}{\pi} \dashint_\R f \,dv \right) \ncr
		&=& u \,G^{-1} [f] - \frac{\ep_I}{\pi |\ep|^2}
			\int_\R f \,dv\,. 
	\eqy
The third term can be dealt with by recognizing that 
	$v F_M$ is only a constant away from $\ep_I$
	and that $H[\ep_I] = \ep_R - 1$, 
	\bq
		G^{-1}\left[v F_M\right] = 
		\frac{T_i}{\pi Z T_e} G^{-1}[\ep_I] = 
		\frac{T_i}{\pi Z T_e} \frac{\ep_I}{|\ep|^2} \,.
	\eq
The sum of these two terms of  \refeq{Gyro} simplify.
	Note that the $\ep's$ depend only on velocity, so they commute 
	with spatial derivatives, 
	\bqy
		\frac{\p}{\p z} G^{-1}[v f] &+& \frac{\p \varphi}{\p z} 
		G^{-1}[v F_M] = u \frac{\p g}{\p z} 
		-
		\frac{\ep_I}{\pi |\ep|^2} \frac{\p}{\p z} \int_\R f \,dv' 
		\ncr
		&+& 
		\frac{Z T_e}{T_i} \frac{\p}{\p z} \int_\R f \,dv'  
		\frac{T_i}{\pi Z T_e} \frac{\ep_I}{|\ep|^2} \ncr
		&=& u \frac{\p g}{\p z}\,.
	\eqy
	
	The nonlinear Poisson bracket term of \eqref{Gyro} involves something that depends only
	on space acting on $f$, thus the $G^{-1}$-transform acts directly on
	$f$. This term also involves $\varphi$, which is proportional to
	the integral of $f$, and we need to express $\varphi$ in terms of $g$.
	The Hilbert transform property \eqref{Hdagger} gives
	\bqy
		\int_\R f \,dv &=& \int_\R G[g] \, dv = 
		\int_\R (\ep_R g + \ep_I H[g]) \, dv
		\ncr
		&=&
		\int_\R (g + H[\ep_I] g + \ep_I H[g]) \, dv \ncr
		&=& \int_\R g \,dv\,.
		 \label{DenInvar}
	\eqy
	Thus we see that the density, or the velocity integral of 
	anything else, is the same in both the original velocity
	coordinates and in the $G^{-1}$-transformed velocity
	coordinates. Also, $\varphi$  remains  unchanged.
	\bq
		G^{-1}\left[\Half \rho_i v_t [\varphi,f]_{x,y} \right] = 
		\Half \rho_i v_t \left[\frac{Z T_e}{T_i}\! \int_\R \!g \, du \, ,
		\, g  \right]_{x,y}\!\!\!.
	\eq
	
	We use the Fokker-Planck collision operator on the right-hand
	side of \eqref{Gyro}. This is convenient because its  interaction with the $G^{-1}$-transform has already been 
	determined in Sec.~\ref{sec:Collisions}. As before, after $G^{-1}$-transforming,
	we get the Fokker-Planck collision operator plus the shielding
	term, and the  shielding term is small and can be neglected.
	
	The source term, $\chi$, is some particular function of velocity.
	After $G^{-1}$-transforming it becomes a new particular function 
	of $u$. Call it $\bar{\chi}(u)$. 
	
	The $G^{-1}$-transformed drift-kinetic equations are: 
	\bqy
	\label{GGyro}
		&&\frac{\p g}{\p t} + u \frac{\p g}{\p z} + \Half \rho_i v_t
		\frac{Z T_e}{T_i} \left[\int_\R g \, du \, , \, g \, \right]  \\
		&& \hspace{ 4cm} =
		\calc[g] + \cals[g] + \bar{\chi}(u) \,.
		\nonumber 
	\eqy
	
	Equation \eqref{GGyro} is a main result of this paper.
	We have shown that if the source and collisions are 
	neglected, the parallel electric field can be exactly 
	eliminated. Also, in the case of collisions, the 
	shielding term is small, so one can eliminate the 
	electric field and retain the use of
	$\calc$ alone.  Thus one can solve a simpler equation
	in $u$ and then $G$-transform back to the original velocity
	coordinates after the calculation of the dynamics is
	finished.
	
	This calculation is typically done using Hermite polynomials
	to discretize velocity space. We include an explicit
	calculation of the $G^{-1}$-transform of the Hermite 
	polynomials in Appendix~\ref{sec:Hermite}.
	
	One of the problems in gyro-/drift-kinetics is the relative
	importance of Landau damping and dissipation due to
	turbulence in the directions perpendicular to the magnetic
	field. Landau damping dominates when the parallel streaming
	time is large compared to the nonlinear correlation
	time.\cite{HatJen13,HatJen14} Otherwise, the perpendicular
	turbulence suppresses Landau damping. Landau damping is
	suppressed when the perpendicular nonlinearity creates
	structures in velocity space which anti-phase-mix (like 
	the plasma echo) and return energy from  fine velocity 
	scales to the spatial dependence of the distribution
	function and fields.\cite{SchDor16} In future work,
	we hope to use the $G$-transform to further illuminate this
	gyrokinetic behavior.
	
	\subsection{$u$-Fluid Equations}
	\label{sec:uFluid}
	
	One technique for solving gyro-/drift-kinetics is to take moments in 
	velocity space. The resulting hierarchy of equations are
	known as the gyrofluid equations.  More specifically,
	define the zeroth, first, second, and third moments of the
	gyrokinetic distribution fluid, respectively, 
	\bqy
		\rho := \int_\R f \,dv \ &,& \
		j := \int_\R v \,f \,dv \ , \ncr
		P := \int_\R v^2 \,f \,dv \ &,& \
		Q := \int_\R v^3 \,f \,dv \ , 
		\label{MomentDefs}
	\eqy
	then take the corresponding moments of \refeq{Gyro} to
	obtain fluid equations. 
	
	
The zeroth moment of \refeq{Gyro} gives  
	\bq
		\frac{\p \rho}{\p t} + \frac{\p j}{\p z} =
		-\frac{\rho_i v_t}{2 L_n} \frac{\p \varphi}{\p y} \,,
		\label{0vmom}
	\eq
%
%
where the moment of the  third term of \eqref{Gyro} is zero since $v F_M$ is an odd function of velocity.
%
 The nonlinear  term is zero since it reduces to the perpendicular Poisson bracket of something with itself. 
%
	The collision operator is a total derivative and it decays as 
	$v \rightarrow \infty$, so its integral is zero. 
%
	The source term is an explicit function of velocity which can 
	be evaluated.
%

	
The first order moment of \eqref{Gyro} is  
	\bq
	 \label{Ohm}
		\frac{\p j}{\p t} + \frac{\p P}{\p z} + \frac{v_t^2}{2} 
		\frac{\p \varphi}{\p z} + \frac{\rho_i v_t}{2} \frac{Z T_e}{T_i} 
		[\rho,j]_{x,y} = -\nu j\,. 
	\eq
%
%
%
	Note, the third term  of \eqref{Ohm} is an explicit function of velocity that can be 
	evaluated
%
	and the nonlinear term reduces to the perpendicular 
	Poisson bracket of the zeroth and first moments. 
%
	From integration by parts, the collision operator reduces to linear drag, 
%
	while the  source term, being  an even function of velocity, vanishes.

	Notice that if all of the fluid variables are uniform in space and
	time,  \eqref{Ohm}  reduces to Ohm's Law: 
	$E = - \p \varphi / \p z \propto j$.
	
	
	Proceeding,  the second order moment of \refeq{Gyro} is 
	\bqy
		&&\frac{\p P}{\p t} + \frac{\p Q}{\p z} + \frac{\rho_i v_t}{2}
		\frac{Z T_e}{T_i} [\rho,P]_{x,y} = -2 \nu \left(P - 
		\frac{v_t^2}{2} \rho \right)\ncr
		&&\hspace{2cm} - \frac{\rho_i v_t^3}{4} \left(
		\frac{1}{L_n} + \frac{1}{L_T} \right) \frac{\p \varphi}{\p y}\,.
	\eqy
%
%
%
	Note, the  moment of the third term  of \eqref{Gyro} 
	vanishes since $v^3 F_M$ is an odd function of velocity, 
%
	while the  nonlinear term reduces to the perpendicular
	Poisson bracket	of the zeroth and second moments. 
%
	With integration by parts,  the collision operator becomes the first term on the right-hand side, and 
%
the source term is an explicit function of velocity that can  be evaluated.
%

	We could continue taking higher moments, with the usual hierarchy where the $n^{th}$ equation 
	is coupled to the $(n+1)^{th}$ moment through the advection term.
	To close the fluid equations, we have to make additional assumptions,
	for example, we could assume that the third moment is some specified
	function of the lower moments: $Q = Q(\rho, j, P)$.
	
	Alternatively, we can write fluid equations in the transformed
	velocity coordinate. These we call the $u$-fluid equations.   That is, 
	instead of taking the velocity of moments of
	\refeq{Gyro}, we  take   $u$ moments of \refeq{GGyro}.
	
	In \refeq{DenInvar}, we saw that the zeroth moments
	in $u$ and $v$ of any function are the same.
	We will now prove similar results for the first and 
	second moments of any function.
	
	For the first moment, consider  
	\bq 
		\int_\R u \, G[g] \, du = 
		\int_\R (u \, \ep_R \, g + u \, \ep_I \, H[g]) \, du\,.
		\label{ID2}
	\eq
	Focusing  on the second term on the right-hand side of \eqref{ID2}, we apply the Hilbert transform property
	\eqref{Hdagger} to move the Hilbert transform off of $g$.
	Then we apply the 
	Hilbert transform property \eqref{Hu} to separate $u$ 
	from $\ep_I$. The correction term is zero since $\ep_I$ is 
	a total derivative. Therefore, 
	\bqy
		\int_\R u \, \ep_I \, H[g] \, du &=& - \int_\R H[u \, \ep_I] \, g \, du
		\ncr
		&=& - \int_\R \! u \, g \, H[\ep_I]\, du - \frac{1}{\pi}\! \int_\R g \, du
		\!\int_ \R \!\ep_I \, du' \ncr
		&=& - \int_\R u \, g \, H[\ep_I]\, du\,, 
	\eqy
	and 
	\bqy
		\int_\R u \, G[g] \, du &=& 
		\int_\R (u \,(1+H[\ep_I]) \, g - u \, g \, H[\ep_I]) \, du
		\ncr
		&=& 
		\int_\R u \, g \, du
		\label{1ID} \,.
	\eqy
	Now let $f = G^{-1}[g]$ in \eqref{1ID}, yielding 
	\bqy
		\int_\R u \, G^{-1}[f] \, du &=& \int_\R \!u \, g \, du
		\ncr
		& =& 
		\int_\R \!u \, G[g] \,du = \int_\R \!u \, f \, du\,. 
		 \label{Mom1Invar}
	\eqy
	Therefore, the  first moment of any function is invariant under $G^{-1}$
	transforms.
	
	For the second moment, we will prove something similar 
	to \refeq{Mom1Invar}, by considering  
	\bq
		\int_\R u^2 \, G[g] \, du = \int_\R (u^2 \, \ep_R \, g + 
		u^2 \, \ep_I \, H[g]) \, du\,.
		\label{ID3}
	\eq
	First focus on the second term of \eqref{ID3}.  We apply  property
	\eqref{Hdagger} to move the Hilbert transform off of $g$, then property \eqref{Hu} twice to separate $u$ 
	from $\ep_I$, yielding  
	\bqy
		\int_\R \!u^2 \, \ep_I \, H[g] \, du &=& 
		- \int_\R g \, H[u^2 \ep_I] \, du
		 \ncr
		&=& - \int_\R g \, u \, H[u \ep_I] \, du\ncr
		&&\hspace{1cm} - \frac{1}{\pi}
		\int_\R g \,du \int_\R u' \, \ep_I \, du' \ncr
		&=& - \int_\R g \, u^2 \, H[\ep_I] \, du 
		\label{ID4}\\
		&&\hspace{.5cm} - \frac{1}{\pi}
		\int_\R u \, g \, du \int_\R \ep_I \, du' 
		\ncr
		&&\hspace{1cm}  - \frac{1}{\pi}
		\int_\R g \, du \int_\R u' \, \ep_I \, du'\,.
		\nonumber
	\eqy
	The second term of the above is zero since $\ep_I$ is a total derivative.
	The integral of $\ep_I$ in the last term can be explicitly evaluated
	for any definition of $\ep_I$: \refeq{EpsDefinition},
	\refeq{EpsDefinitionMod}, or \refeq{EpsDefGyro}. We focus on
	\refeq{EpsDefinitionMod} and find  
	\bqy
		\frac{1}{\pi} \int_\R u' \, \ep_I \, du' &=& 
		\frac{1}{\pi} \int_\R u' \,  \pi \frac{Z T_e}{T_i} u' F_M \, du' 
		\ncr
		&=& 
		\frac{Z T_e}{T_i} \int_\R (u')^2 F_M \, du' 
		=
		\frac{Z T_e}{T_i} \frac{v_t^2}{2}\,.
		\nonumber
	\eqy
	Once again, let $g = G^{-1}[f]$. Recall that the 
	zeroth moment of any function is invariant under $G^{-1}$,
	so 
	\bqy
		\int_\R u^2 \, G[g] \, du &=& \int_\R u^2 \,\ep_R \, g \, du
		\ncr
		&&
		- \int_\R u^2 \,H[\ep_I] \, g \, du 
		- \frac{Z T_e}{T_i} \frac{v_t^2}{2} \!\int_\R g \, du \ncr
		&=& \int_\R u^2 \, g \, du - \frac{Z T_e}{T_i} \frac{v_t^2}{2} 
		\int_\R g \, du\,,
		 \label{GMom2} \\
		\int_\R u^2 \, G^{-1}[f] \, du &=& 
		\int_\R u^2 \, f \, du + \frac{Z T_e}{T_i} \frac{v_t^2}{2} 
		\int_\R f \, du \,.
		\label{G1Mom2}
	\eqy
	The transformed second moment is equal to the original 
	second moment plus a correction proportional to the 
	zeroth moment. If we continue this pattern, we find that
	each transformed moment is equal to the original moment
	plus some corrections proportional to lower order moments
	with the same parity.
	
	Define the second and third $u$ moments as:
	\bq
		\bar{P} := \int_\R u^2 \, g \, du \ , \
		\bar{Q} := \int_\R u^3 \, g \, du  \,.
		\label{uMomentDefs}
	\eq
	Note that because of \refeq{DenInvar} and
	\refeq{Mom1Invar}, the zeroth and first moments
	remain unchanged. The second $u$ moment is related to 
	the original moment according to
	\bq
		\bar{P} = P + \frac{Z T_e}{T_i} \frac{v_t^2}{2} \rho \,.
		\label{PresRel}
	\eq
	This same relationship holds between the two pressures (as opposed to the second moments) 
	because the difference between the second moment and the
	pressure depends only on the first and zeroth moments, 
	both of which are invariant under $G$-transforms.
	
	Beginning with the zeroth moment of \refeq{GGyro},
	the first two terms are simple, as before. 
	The nonlinear term is zero since it reduces to 
	the perpendicular Poisson bracket of something with itself.
	Using the argument \refeq{DenInvar}, the zeroth $u$ moment
	of the right-hand side must be the same as the
	zeroth $v$ moment. The collision operator's contribution 
	and the source term's contribution are the same as before.
	We get exactly \refeq{0vmom} again.
	
	We next consider the first order moment of \refeq{GGyro}.
	The first term and second terms are obvious, while
	the nonlinear term reduces to the perpendicular Poisson
	bracket of the zeroth and first moment.
	
	When dealing with the source term,  it will be helpful to
	utilize parity. We note that typically the equilibrium
	distribution function is even, derivatives and Hilbert
	transforms by \eqref{HParity} reverse parity, 
	$\ep_I \propto \p f_0 / \p v$ (or $\ep_I \propto v F_M$)
	is  odd, and  $\ep_R = 1 + H[\ep_I]$ is even. Thus the
	$G$-transform and its inverse also preserve parity.	
	Because the source term of \refeq{GyroSource} in the
	original coordinates is even, the $G^{-1}$-transformed
	source term $\bar{\chi}(u)$ is also even. The first moment
	of any even function is zero, so we get no contribution from
	the source.  
	
	Proceeding to the collision and shielding terms, we can
	deal with them using \refeq{Mom1Invar}: 
	\bqy
		\int_\R u (\calc[g] + \cals[g]) \, du &=&
		\int_\R u \, G^{-1} [\calc[ G [g]]] \, du 
		=
		\int_\R \!u \, \calc[ G [g]] \, du \ncr
		&=& \nu \int_\R \!u  \frac{\p}{\p u} \!\left(\!\frac{v_t^2}{2}
		\frac{\p}{\p u} G[g] + u \, G[g]\! \right) \! du \ncr
		&=& - \nu \! \int_\R \!\left( \frac{v_t^2}{2} \frac{\p}{\p u} G[g]
		+ u \,G[g] \right)  du \ncr
		&=& - \nu \int_\R u \,G[g] \, du = 
		-\nu \int_\R u \,g \,du \ncr
		&=& - \nu j \,.
	\eqy
	Putting it all together, we find the first moment \eqref{GGyro} is
	\bq
		\frac{\p j}{\p t} + \frac{\p \bar{P}}{\p z} + 
		\Half \rho_i v_t \frac{Z T_e}{T_i} \,[\rho,j]_{x,y}
		= - \nu j\,.
	\eq
	
	We look for something corresponding to Ohm's law in vain. 
	The only way that all of the $u$ fluid quantities can be uniform
	in space and time is if $j$ is zero. The $G^{-1}$-transform
	removes the electric field, so there is nothing to balance 
	the equilibrium $u$ current against.
	
	Finally we consider the second order moment of \refeq{GGyro}.
	The first and second terms are again trivial and the nonlinear terms is again a Poisson bracket. 
	
	Equation \refeq{G1Mom2} allows us to determine the second moment 
	of the $G^{-1}$-transformed source term:
	\bqy
		\int_\R u^2 \, G^{-1}[\chi] \, du  &=&
		\int_\R u^2 \, \chi \, du + \frac{Z T_e}{T_i} \frac{v_t^2}{2}
		\int_\R \chi \, du \ncr
		&=& -\frac{\rho_i v_t^3}{4} \frac{\p \varphi}{\p y} 
		\left(\frac{1}{L_n} \left(1 + \frac{Z T_e}{T_i}\right) 
		+ \frac{1}{L_T}\right)\,.
		\nonumber
	\eqy
	
	We can now use \refeq{G1Mom2} and then \refeq{DenInvar} and
	\refeq{GMom2} to deal with the collision and shielding terms
	simultaneously.
	Recall that the collision operator is a total derivative, so
	the integral of it acting on anything over all space is zero.  Thus, 
	\bqy
		\int_\R u^2  (\calc[g] &+& \cals[g]) \, du = 
		\int_\R u^2 \, G^{-1}[\calc[G[g]]] \, du \ncr
		&=& \int_\R u^2 \,\calc[G[g]] \, du
		+ \frac{Z T_e}{T_i} \frac{v_t^2}{2}	\int_\R \calc[G[g]] \, du \ncr
		&=& \nu \int_\R u^2 \, \frac{\p}{\p u} \left( \frac{v_t^2}{2} 
		\frac{\p}{\p u} G[g] + u \, G[g] \right) du \ncr
		&=& \nu v_t^2 \int_\R G[g] \, du - 2\nu \int_\R u^2 \, G[g] \, du\ncr
		&=& \nu v_t^2 \int_\R g \, du - 2\nu \int_\R u^2 \,g \, du 
		\ncr
		&&\hspace{2cm} + 2\nu \frac{Z T_e}{T_i} \frac{v_t^2}{2} \int_\R g \, du \ncr
		&=& \nu v_t^2 \left(1 + \frac{Z T_e}{T_i} \right) \rho
		- 2 \nu \bar{P}\,.
	\eqy
	
	Finally, the second moment of \eqref{GGyro} is 
	\bqy
		&&\frac{\p \bar{P}}{\p t} + \frac{\p \bar{Q}}{\p z} + 
		\Half \rho_i v_t \,\frac{Z T_e}{T_i} \,[\rho, \bar{P}]_{x,y}
		= 
		\ncr
		&&\hspace{1cm}  - 2 \nu \left(\bar{P} - \frac{v_t^2}{2} \left(1 + 
		\frac{Z T_e}{T_i} \right) \rho \right) 
		\ncr
		&&\hspace{1.5cm}
		-\frac{\rho_i v_t^3}{4} \frac{\p \varphi}{\p y} 
		\left(\frac{1}{L_n} \left(1 + \frac{Z T_e}{T_i}\right) 
		+ \frac{1}{L_T}\right)\,.
		\nonumber 
	\eqy
	
	Once again, we get a hierarchy of coupled equations.
	These equations could be used as an alternative to 
	gyrofluid equations. They are simpler because we have
	eliminated the parallel electric field. 
	
	The challenge is that we do not 
	intuitively understand what the $u$ moments mean. 
	We cannot transform back to the usual gyrofluid variables
	because we have already integrated over the velocity.
	This also makes it more difficult to determine an
	appropriate closure. Some closures are similar. 
	If a barotropic closure is appropriate for the original
	moments, then a barotropic closure is also appropriate 
	for the $u$ moments because the difference between the 
	two pressures only depends on the density \refeq{PresRel}.
	However, the $u$ pressure would not be expected to have a
	single polytropic index, even if the original pressure does.  Also, other closure ideas may apply (e.g.\ those of Ref.~\onlinecite{pjmPCT14}). 
	Once we have developed some intuition
	about dynamics in $u$ space from gyrokinetic models, 
	we will be better able to interpret and use these 
	new gyrofluid equations.

\section{Conclusion}
\label{sec:Conclusion}
	
	We described how the  one dimensional linearized Vlasov-Poisson system can be
	exactly solved using the $G$-transform, an integral transform based on the Hilbert
	transform,  that removes the electric field term. 
	 In terms of this integral transform, Landau damping appears as the Riemann-Lebesgue Lemma: 
	a rapidly oscillating function integrates to zero.
	
	The $G$-transform can be used for any kinetic theory
	with one velocity dimension. Given that efficient Hilbert transform  algorithms exist, it 
	is numerically easy to implement.

	We analyzed how the $G$-transform interacts with the Fokker-Planck
	collision operator. The commutator between the $G$-transform and
	the collision operator gives rise to an additional term, which 
	we call the shielding term. The shielding term was shown to   be small. 
	If there is no small scale structure in velocity space, then 
	the collision term and the shielding term are small since they
	are multiplied by a small parameter, the collision frequency. 
	If there is small scale
	structure in velocity space, then the collision term is significant
	since it contains the highest order velocity derivative. The
	shielding term is still unimportant since its velocity derivatives
	are all lower order than the collision term.
	
	If we drop the shielding term, then the resulting
	advection-diffusion equation can be exactly solved. We wrote an
	explicit solution for simple initial conditions and used it
	to determine when advection dominates and when collisions dominate
	the equation (\refFig{fig:Coll2Adv}). We then numerically solved 
	the advection-diffusion equation for more realistic initial 
	conditions and showed that our conclusions are not substantially
	different from the simple initial conditions 
	(\refFig{fig:CompTerms}). We then discussed how the shielding
	term could be included as a perturbation.
	
	Although we focused on the Fokker-Planck collision operator,
	similar arguments also apply for any other collision operators that
	are local in $v$. The shielding term will be small because it is
	multiplied by a small parameter and has only lower order velocity
	derivatives than the collision term. If future work, we intend
	to extend this argument to more complicated collision operators, 
	such as the one dimensional linearized Landau-Boltzmann operator
	\cite{Bol72,LanLif10} and pitch angle scattering. \cite{HasWib68}
	
	The most fruitful applications of the $G$-transform will
	likely be found in gyro-/drift-kinetics, which have one velocity
	dimension while still capturing  much of the relevant physics 
	for tokamak and space plasmas.  For this system the $G$-transform is
	slightly modified: since the Poisson
	equation has been replaced by quasineutrality, the
	$G$-transform no longer has spatial dependence. It
	removes the linear electric field term, leaves the
	nonlinearity unchanged, and only modifies the collision
	operator by a small	perturbation \refeq{GGyro}.
	
	We compared gyrofluid equations found by taking moments in
	both the original velocity space and in the transformed $u$
	space. There is nothing corresponding to Ohm's law in the 
	transformed gyrofluid equations: the electric field has
	been removed, so the only things that can balance the first
	$u$ moment are spatial and time derivatives of other $u$ moments.
	
	Gyrokinetic codes often use Hermite polynomials times
	a Gaussian as a basis in velocity space. We give an
	explicit expression for the transformed basis elements
	\refeq{ExplicitGHermite}.
	
	We are currently working to include the $G$-transform in
	gyrokinetic codes. Numerically, computing the $G$-transform
	involves computing a Hilbert transform and storing the real
	and imaginary parts of the plasma dielectric function.
	Since the transform removes the linear electric field, it
	removes the coupling between each Hermite polynomial and
	the zeroth Hermite polynomial. Hermite polynomials
	are only coupled to their neighbors through the advection 
	term. The perpendicular nonlinearity coupling of different
	perpendicular wave numbers remains unchanged.
	
	One of the interesting features of gyrokinetics is that 
	the importance of Landau damping to the dissipation rate
	can change dramatically depending on the nonlinearity.
	We hope to use the $G$-transform to inform a model with
	only a few modes that shows the same behavior.


\section*{Acknowledgment}
\noindent   Supported by U.S. Dept.\ of Energy Contract \# DE-FG05-80ET-53088.





\bibliographystyle{apsrev}

\bibliography{jmh}
  

\appendix

\section{Comparing Collision Operators}
\label{sec:app:CompColl}

After $G$-transforming the Fokker-Planck collision operator,
we get the same collision operator plus some corrections 
which we call the shielding terms. The shielding terms are
expected to be small because they are multiplied by the 
small parameter $\nu$ and do not contain the highest 
order velocity derivative. Two of the terms in the collision 
operator are also the same order of magnitude of the 
shielding operator. We have a choice between dropping 
everything but the second velocity derivative term or just
dropping the shielding term. Both equations can be 
solved analytically for general initial conditions. 

If we drop everything except the second velocity derivative, 
the resulting equations, 
\bq
	\frac{\p g_k}{\p t} + i k u \,g_k =
	\nu \frac{v_t^2}{2} \frac{\p^2 g_k}{\p u^2} \,,
\eq
with initial conditions $\initcond{g_k}(u)$, has solution:
\bqy
	g_k(t,u)&=& \frac{1}{\sqrt{\pi} v_t \sqrt{2 \nu t}} \
	\int du' \,\initcond{g_k}(u') \exp\Big[
	-\frac{i}{2} k (u+u') t \ncr 
	& & - \frac{(u-u')^2}
	{2 v_t^2 \nu t} -\frac{1}{24} k^2 v_t^2 \nu t^3\Big]
	\label{SecDerivGenSol}
\eqy
When we plug in Gaussian initial conditions
\refeq{GaussIC}, we get the 
solution analyzed before \refeq{SolNoShield}.

If we drop the shielding terms, but keep the entire collision
operator, the resulting equations,
\bq
	\frac{\p g_k}{\p t} + i k u \,g_k \ = \ 
	\nu \left(\frac{v_t^2}{2} \frac{\p^2 g_k}{\p u^2} + 
	u \frac{\p g_k}{\p u} + g_k\right) \,,
	\label{AdvDiffEntireColl}
\eq
with initial conditions $\initcond{g_k}(u)$, has solution:
\bqy
	g_k(t,u) &=& \frac{1}{\sqrt{\pi} v_t \sqrt{1-e^{-2\nu t}}}
	\int du' \,\initcond{g_k}(u') \ncr
	& & \exp\Big[- i \frac{k}{\nu} 
	(u+u') \frac{1-e^{-\nu t}}{1+e^{-\nu t}}
	- \frac{(u - u'e^{-\nu t})^2}{v_t^2 (1-e^{-2\nu t})} \ncr
	& & + \frac{k^2 v_t^2}{\nu^2} \Big(\frac{1-e^{-\nu t}}
	{1+e^{-\nu t}} - \frac{\nu t}{2}\Big)\Big]
	\label{FullCollGenSol}
\eqy

To compare these solutions, expand \refeq{FullCollGenSol} 
for small $\nu t$. 
\bqy
	1-e^{-2\nu t} &\approx& 2 \nu t \\
	\frac{1-e^{\nu t}}{1+e^{\nu t}} &=& 
	\tanh\left[\frac{\nu t}{2}\right] \ \approx \ 
	\frac{\nu t}{2} - \frac{\nu^3 t^3}{24}
\eqy
When we substitute these in and drop anything higher than 
first order in $\nu$, we find that \refeq{FullCollGenSol} agrees
with \refeq{SecDerivGenSol}.

We know that the solutions will decay 
at a rate given by \refeq{DecayTime}, which is much shorter
than $1/\nu$. By the time $\nu t$ gets close to one, the 
perturbation will have already decayed to close to zero.

Some of the terms we drop also contain velocity. For example, 
we drop $ {i}  k u \nu^2 t^3/24 $. This
term could become large at sufficiently large velocities:
\bq
	u \gtrsim \frac{24}{k \nu^2 t^3}
\eq
Since $\nu$ is a small parameter, this will be large compared
to $v_t$ until well past the decay time. Since any reasonable
initial condition decays at large velocity, the distribution
function here will be negligibly small.

The two solutions are almost equal 
whenever the distribution function is significantly 
different from zero. Since the shielding terms are the same
order of magnitude as the difference between these two collision
operators, this analysis gives further indication of the 
legitimacy of dropping them.

	\section{Hermite Polynomials}
	\label{sec:Hermite}
	
	Many gyrokinetic codes use Hermite polynomials multiplied
	by a Gaussian as a basis for velocity space. To use the 
	$G$-transform to simplify gyrokinetic codes, we will
	need to know how to $G^{-1}$-transform Hermite polynomials.
	
	We define the Hermite polynomials as:
	\bq \label{HermiteDef}
		H_n(\zeta) := 
		\frac{(-1)^n}{\sqrt{n! \,2^n \sqrt{\pi}}} \,e^{\zeta^2}
		\left(\frac{d}{d\zeta}\right)^n e^{-\zeta^2} \,.
	\eq
	There are multiple normalizations used for the Hermite polynomials.
	Changing the normalization does not substantially change our 
	results.
	
	The derivatives of the Gaussian can be written explicitly using
	the Hermite	polynomials. Define the constants out front to be $a_n$:
	\bqy	
		\left(\frac{d}{d\zeta}\right)^n e^{-\zeta^2} &=& 
		(-1)^n \sqrt{n! \,2^n \sqrt{\pi}} \,H_n(\zeta) \,e^{-\zeta^2}
		\ncr
		&=:& a_n H_n(\zeta) \,e^{-\zeta^2} \,. \label{ADef}
	\eqy
	Note that $a_n = - \sqrt{2 n} \,a_{n-1}$.
	
	There is a simple expression for the derivative of a Hermite polynomial:
	\bqy
		\frac{d}{d \zeta} H_n(\zeta) 
		&=& \sqrt{2n} \,H_{n-1}(\zeta) \,.
	\eqy
	
	Iterate this to get the expression for an arbitrary derivative of 
	a Hermite polynomial. For any integers $n,m$ with $m \leq n$, 
	\bq \label{HermDerivN}
		\left(\frac{d}{d \zeta}\right)^m H_n(\zeta) = 
		\sqrt{\frac{2^m \,n!}{(n-m)!}} \,H_{n-m}(\zeta) \,.
	\eq

	There is an explicit recurrence relation for Hermite polynomials:
	\bqy
		a_{n+1} H_{n+1} 
		&=& - 2 n \,a_{n-1} H_{n-1} - 2 \zeta \,a_n H_n \,.
	\eqy
	
	$G^{-1}$-transform the Hermite polynomials, multiplied by a 
	Gaussian. Use Hilbert transform property \refeq{HDeriv}.
	Recognize the plasma Z function \refeq{ZDef} when it appears:
	\bqy
		G^{-1}\left[\frac{1}{a_n} \frac{d^n}{d\zeta^n}
		e^{-\zeta^2}\right] 
		&=& \frac{1}{a_n} \bigg( \frac{\ep_R}{|\ep|^2}
		\frac{d^n}{d\zeta^n} e^{-\zeta^2}\ncr
		&&\qquad  - \frac{\ep_I}{|\ep|^2}
		H\left[\frac{d^n}{d\zeta^n} e^{-\zeta^2}\right] \bigg)\ncr
		&=& \frac{1}{a_n} \bigg(\frac{\ep_R}{|\ep|^2}
		\frac{d^n}{d\zeta^n} e^{-\zeta^2}  \label{GInvHPoly1}
		\\
		&&\qquad  - \frac{\ep_I}{\sqrt{\pi}
		|\ep|^2} \frac{d^n}{d\zeta^n} Z(\zeta) \bigg) \,.
		\nonumber
	\eqy
	
	The derivative of the plasma Z function is given by \refeq{ZDeriv}.
	The expression for an arbitrary derivative of the plasma Z function
	can be proved by induction:
	\bqy
		\frac{d^n Z}{d \zeta^n} &=& a_n H_n(\zeta) \,Z(\zeta) - 2
		\sum\limits_{j=0}^{n-1} a_j \!\left(\frac{d}{d\zeta}
		\right)^{n-1-j}\hspace*{-.8cm} H_j(\zeta) \,.
		 \label{ZDerivN}
	\eqy
%
%
%
	Combine \refeq{ZDerivN} with \refeq{HermDerivN} to get an 
	expression for the derivatives of the plasma Z function 
	explicitly in terms of Hermite polynomials.
	Plug in the explicit expression for $a_j$ \refeq{ADef}.
	Note that the derivatives in the sum will only be nonzero if
	$n-1-j \leq j$, i.e. $j \geq \frac{n-1}{2}$:
	\bqy
		\frac{d^n Z}{d \zeta^n} 
		&=& a_n H_n(\zeta) \,Z(\zeta) \label{ExplicitZDerivN}\\
		&&\qquad 
		- 2 \sum_{j \geq \frac{n-1}{2}}
		^{n-1} (-1)^j \,j! \,\sqrt{\frac{2^{n-1}}{(2j-n+1)!}} \
		H_j(\zeta)\,.
		 \nonumber
	\eqy
	
	We can now get an expression for the $G^{-1}$-transform of
	the Hermite polynomials in terms of a Gaussian, the plasma Z
	function, and Hermite polynomials in the new velocity coordinate:
	\bqy
		&&G^{-1} \left[\frac{1}{a_n} \,\frac{d^n}{d\zeta^n}
		e^{-\zeta^2}\right] 
		= \frac{\ep_R}{|\ep|^2} \,H_n(\zeta) \,e^{-\zeta^2}
		\ncr
		&&  \hspace{2cm} -  \frac{\ep_I}{\sqrt{\pi} |\ep|^2} \,H_n(\zeta) \,Z(\zeta)
		\ncr
		&& \hspace{2cm} + \frac{2}{\sqrt{\pi}} \frac{\ep_I}{|\ep|^2} 
		\sum_{j \geq \frac{n-1}{2}}^{n-1} 
		\frac{(-1)^j \,j!}{(-1)^n\sqrt{n! \,2^n \sqrt{\pi}}}
		\ncr
		&&
		\hspace{3.0cm} \times \sqrt{\frac{2^{n-1}}{(2j-n+1)!}} \,H_j(\zeta)\,. 
		\label{ExplicitGHermite}
	\eqy
	Note that the $G^{-1}$-transform of the Hermite polynomials only involves
	other Hermite polynomials of lower order, so it will not cause any
	problems when you truncate the Hermite series.

\end{document}